\documentclass{aa}  
\usepackage{natbib}
\bibpunct{(}{)}{;}{a}{}{,} 
\usepackage{graphicx}
\usepackage{txfonts}
\newcommand{\vap}{v_{\mathrm{Ap}}}

\newcommand{\ta}{\tau_{\mathrm{Ap}}}

\newcommand{\pd}{\partial}

\newcommand{\ck}{v_{\rm k}}
\newcommand{\ckp}{v_{\rm kp}}
\newcommand{\cke}{v_{\rm ke}}

\newcommand{\rp}{\rho_{\rm p}}
\newcommand{\re}{\rho_{\rm e}}
\newcommand{\rc}{\rho_{\rm c}}
\newcommand{\lp}{L_{\rm p}}

\begin{document}

	\title{Kink oscillations of flowing threads in solar prominences}

	\titlerunning{Kink oscillations of flowing prominence threads}

   \author{R. Soler\inst{\ref{leuven},\ref{uib}} \and M. Goossens\inst{\ref{leuven}}}
\offprints{R. Soler}
\institute{Centre for Plasma Astrophysics, Department of Mathematics, Katholieke Universiteit Leuven,
              Celestijnenlaan 200B, 3001 Leuven, Belgium \\
                       \email{roberto.soler@wis.kuleuven.be}
 \label{leuven}
 \and
 Solar Physics Group, Departament de F\'isica, Universitat de les Illes Balears,
             E-07122 Palma de Mallorca, Spain \label{uib} }

 	 \date{Received XXX / Accepted XXX}

  \abstract
   {Recent observations by Hinode/SOT show that MHD waves and mass flows are simultaneously present in the fine structure of solar prominences.}
   {We investigate standing kink magnetohydrodynamic (MHD) waves in flowing prominence threads from a  theoretical point of view. We model a prominence fine structure as a cylindrical magnetic tube embedded in the solar corona with its ends line-tied in the photosphere. The magnetic cylinder is composed of a region with dense prominence plasma, which is flowing along the magnetic tube, whereas the rest of the flux tube is occupied by coronal plasma.}
   {We use the WKB approximation to obtain analytical expressions for the period and the amplitude of the fundamental mode as functions of the flow velocity. In addition, we solve the full problem numerically by means of time-dependent simulations.}
   {We find that both the period and the amplitude of the standing MHD waves vary in time as the prominence thread flows along the magnetic structure. The fundamental kink mode is a good description for the time-dependent evolution of the oscillations, and the analytical expressions in the WKB approximation are in agreement with the full numerical results.}
   {The presence of flow modifies the period of the oscillations with respect to the static case. However, for realistic flow velocities this effect might fall within the error bars of the observations. The variation of the amplitude due to the flow leads to apparent damping or amplification of the oscillations, which could modify the real rate of attenuation caused by an additional damping mechanism.}


   \maketitle


\section{Introduction}

Recent observational evidence of ubiquitous periodically varying features in the solar corona \citep[e.g.,][]{tomczyk2007,jess2009,tomczyk2009,wang2009} has raised the debate on whether these observations are caused by magnetohydrodynamic (MHD) waves or by quasi-periodic flows \citep[see, e.g.,][]{depontieu2010}. There seem to be strong theoretical arguments supporting the wave interpretation \citep[e.g.,][]{erdelyi2007,tom2008,TGV,VTG2010}. However, waves and flows are not mutually exclusive and, in fact, both phenomena have been simultaneously observed in the fine structure of solar prominences \citep[e.g.,][]{okamoto}. This offers us the opportunity to study the interaction between waves and flows in the solar atmosphere.

The fine structure of solar prominences is clearly visible in the high-resolution H$\alpha$ and Ca II H-line images from the Solar Optical Telescope (SOT) aboard the Hinode satellite \citep[e.g.,][]{okamoto,berger,chae,ning,brigi,chae2010}. When observed above the limb, vertical structures are commonly seen in quiescent prominences \citep[e.g.,][]{berger,chae,chae2010}, while horizontal threadlike structures are usually observed in active region prominences \citep[e.g.,][]{okamoto}. Although it is apparently difficult to reconcile both pictures, some authors \citep[e.g.,][]{brigi} have suggested that vertical threads might actually be a pile up of horizontal threads which appear as vertical structures when projected on the plane of the sky. This idea is consistent with H$\alpha$ observations of filaments on the solar disk from the Swedish Solar Telescope \citep[e.g.,][]{lin08,lin09}, in which the filament fine structure is seen as thin and long dark ribbons. On the other hand, other authors \citep[e.g.,][]{chae2010} have argued that vertical threads are real and are an indication of the existence of vertical magnetic fields in quiescent prominences. Thus, it remains unclear whether all prominences have the same magnetic structure or, on the contrary, the magnetic field in quiescent prominences is predominantly vertical and active region prominences have horizontal fields. A recent review on the properties of prominence threads can be found in \citet{linrev}.

There are many evidences of transverse oscillations of the fine structures of both active region and quiescent prominences, which have been interpreted in terms of kink MHD waves \citep[see the recent reviews by][]{ballester, oliver, arreguiballester}. The reported periods are usually in a narrow band between 2 and 10 minutes, while the oscillations are typically damped after a few periods. In addition, flows and mass motions in prominences have been also reported \citep[e.g.,][]{zirker98, wang1999, kucera2003, lin2003,ahn2010}. The typical flow velocities are less than 30~km~s$^{-1}$ in quiescent prominences, although larger values up to 40--50~km~s$^{-1}$ have been observed in active region prominences.

The work of \citet{okamoto} is an example of simultaneous transverse oscillations and mass flows in prominence fine structures observed with Hinode/SOT. In the present paper we focus on the theoretical analysis of the event reported by \citet{okamoto}. Similar observations of simultaneous flows and oscillations have been reported by \citet{ofmanwang} in coronal loops and by \citet{cao2010} in filament footpoints. Also, the recent work by \citet{antolin} on observations of transverse oscillations of loops with coronal rain is relevant for our present theoretical investigation.  \citet{okamoto} observed an active region prominence formed by a myriad of horizontal magnetic flux tubes which are partially outlined by threads of cool and dense prominence plasma. The magnetic tubes are probably rooted in the solar photosphere. Although only the part of the tubes filled with prominence material can be seen in the Ca II H-line images, the length of the whole magnetic tube must be much longer than the length of the prominence threads, which is roughly between 3,000~km and 16,000~km.  \citet{okamoto} detected that some threads were flowing along the magnetic tubes and simultaneously oscillating in the vertical direction. The mean period of the oscillations was 3~min and the apparent flow velocity on the plane of sky was around 40~km~s$^{-1}$.  The oscillations were in phase along the whole length of the threads, and the wavelength was estimated to be at least 250,000~km.

The event observed by \citet{okamoto} was  studied from a theoretical point of view by \citet{hinode}, who interpreted the observations in terms of standing kink MHD modes supported by the magnetic structure \citep[see, e.g.,][]{edwinroberts,diaz2002,goossens2009,solerstatic}. An interpretation of the observations by \citet{okamoto} in terms of kink modes was also suggested by \cite{erdelyi2007} and \citet{tom2008}.   \citet{hinode} used the observed wave properties provided by \citet{okamoto} to perform a seismological estimation of a lower bound of the prominence Alfv\'en speed. The time-dependent numerical simulations by \citet{hinode} suggested that the influence of the flow on the period was small. Nevertheless, the precise effect of the flow was not assessed in their work because a detailed parametric study was not performed. The purpose of this paper is to advance the analysis of the event observed by \citet{okamoto} by combining both analytical and numerical methods. In the analytical part, we use the WKB approximation to assess the effect of the flow on the period and the amplitude of the transverse oscillations. Expressions of these quantities as functions of the relevant parameters of the model are obtained. In the numerical part, we go beyond the WKB approximation and solve the full time-dependent problem. The implications of our results for the magneto-seismology of prominences are also discussed.

This paper is organized as follows. Section~\ref{sec:model} contains the model configuration and the basic governing equations. The analytical investigation of standing kink MHD waves in flowing prominence threads using the WKB approximation is included in Section~\ref{sec:analytics}, while the full numerical solution of the time-dependent problem is performed in Section~\ref{sec:numerics}. Finally, our results are discussed in Section~\ref{sec:discussion}.

\section{Model and basic equations}
\label{sec:model}

\begin{figure*}[!htp]
\centering
\includegraphics[width=1.75\columnwidth]{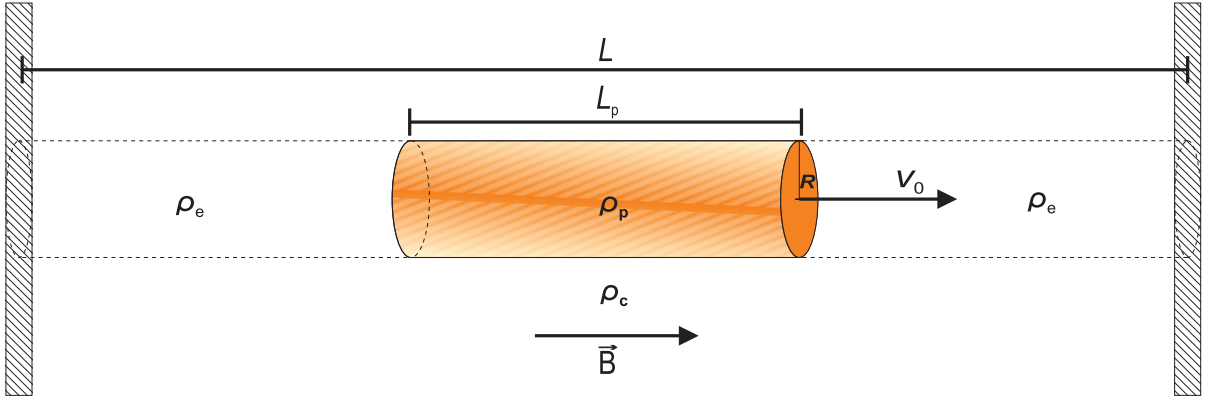}
\caption{Sketch of the prominence fine structure model adopted in this work. \label{fig:model}}
\end{figure*}

The background model in which the waves are superimposed is schematically shown in Figure~\ref{fig:model}. It is composed of a straight and cylindrical magnetic flux tube of radius $R$ and length $L$, whose ends are fixed at two rigid walls representing line-tying at the solar photosphere. The $z$-axis is chosen so that it coincides with the axis of the tube, and the photospheric walls are located at $z= \pm L/2$.  The magnetic tube is partially filled with prominence plasma of density $\rp$, while the rest of the tube, i.e., the evacuated part, is occupied by less dense plasma of density $\re$. The density of the external plasma is  the density of the coronal medium, $\rc$. The length of the prominence region (thread) is $\lp$. The thread flows along the tube as a block with constant speed $v_0$. The magnetic field is ${\bf B} = B \hat{e}_z$, with $B$ homogeneous. As the $\beta=0$ approximation is used in the present work, with $\beta$ the ratio of the gas pressure to the magnetic pressure, the plasma temperature is irrelevant for the study of kink MHD waves supported by the model. In the absence of flow, standing kink MHD waves supported by the present model were investigated by \citet{joarder97} and \citet{diaz2001} in Cartesian geometry, and by \citet{diaz2002}, \citet{dymovaruderman}, \citet{diazperiods}, and \citet{solerstatic} in cylindrical geometry. 

 We adopt the TT approximation, which is valid for $R/L \ll 1$ and $R/\lp \ll 1$. To check whether or not this approximation is reasonable in the context of prominence threads, we take into account that the values of $R$ and $\lp$ reported by the observations \citep[e.g.,][]{lin04,okamoto,lin08} are in the ranges 50~km~$\lesssim R \lesssim$~300~km and 3,000~km~$\lesssim \lp \lesssim$~28,000~km, respectively, and assume $L \sim 10^5$~km as a typical length for the magnetic tube. We obtain $R/\lp$ and $R/L$ in the ranges $3\times 10^{-3} \lesssim R/L_{\rm p} \lesssim 0.1$ and $5\times 10^{-4} \lesssim R/L \lesssim 3\times 10^{-3}$, meaning that the use of the TT approximation is justified in prominence fine structures. In the case $v_0 = 0$, the basic equation governing linear kink MHD waves of the flux tube in the TT approximation was derived by \citet{dymovaruderman} in their Equation~(21).  In the absence of flow, the TT approximation was used by \citet{dymovaruderman}, \citet{diazperiods}, and \citet{solerstatic}. The results of these works fully agree with the general results beyond the TT approximation by \citet{joarder97}, \citet{diaz2001}, and \citet{diaz2002}.  

In the presence of flow, an intuitive generalization of Equation~(21) of \citet{dymovaruderman} was performed by \citet{hinode} in their Equation~(2). Mathematically, \citet{morton2} also considered the variation of  density with time and obtained a similar expression in their Equation~(18). We refer the reader to \citet{hinode} and \citet{morton2} for a detailed derivation of the basic equation. In the mathematical derivation of \citet{morton2} it is assumed that the difference of the flow velocity between the internal and external plasma is small, i.e., much smaller than the Alfv\'en velocity. So, we restrict our present investigation to values of the flow velocity that satisfy $v_0 / \vap \ll 1$, where $\vap = \frac{B}{\sqrt{\mu \rp}}$ is the prominence Alfv\'en speed. Assuming $B=50$~G and $\rp=10^{-10}$~kg~m$^{-3}$ as realistic values of the magnetic field strength and density in active region prominences, we obtain $\vap \approx 446$~km~s$^{-1}$. Since the flow velocities on the plane of sky estimated by \citet{okamoto} are in the interval between 15~km~$^{-1}$ to 46~km~s$^{-1}$ (see their Table~1), the restriction $v_0 / \vap \ll 1$ is satisfied for realistic parameters in prominences.

Thus, the governing equation we study in the present work is
\begin{equation}
 \frac{\pd^2 v_r(z,t)}{\pd t^2} - \ck^2(z,t) \frac{\pd^2 v_r(z,t)}{\pd z^2} = 0, \label{eq:kink}
\end{equation}
which has to be solved along with the condition of line-tying at the photosphere expressed as $v_r(\pm L / 2,t) = 0$, and a given initial condition at $t=0$. In Equation~(\ref{eq:kink}), $v_r(z,t)$ is the radial velocity perturbation at the tube boundary and $\ck(z,t)$ is the kink speed, which in our model is a function of $z$ and $t$, namely
\begin{equation}
 \ck(z,t) = \left\{ \begin{array}{lll}
                     \ckp & \textrm{if} & \left| z -z_0 - v_0 t \right| \leq \lp/2, \\
		      \cke & \textrm{if} & \left| z -z_0 - v_0 t \right| > \lp/2,
                    \end{array} \right. \label{eq:kinkspeed}
\end{equation}
 where
\begin{equation}
 \ckp = \sqrt{\frac{2 B^2}{\mu \left( \rp + \rc \right)}}, \qquad \cke = \sqrt{\frac{2 B^2}{\mu \left( \re + \rc \right)}},
\end{equation}
with $\mu$ the magnetic permittivity, and $z_0$ corresponds to the position of the center of the prominence thread with respect to the center of the magnetic tube at $t=0$. Note that $z_0 < 0$ if the thread is initially located on the left-hand side of the center of the tube, whereas $z_0 > 0$ if the thread is initially located on the right-hand side. We see that the flow does not explicitly appear in Equation~(\ref{eq:kink}) since it is enclosed in the definition of $\ck(z,t)$ given in Equation~(\ref{eq:kinkspeed}). \citet{hinode} performed time-dependent simulations and solved Equation~(\ref{eq:kink}) numerically. Here, our aim is to solve Equation~(\ref{eq:kink}) by using both analytical and numerical methods. 

\section{Analytical investigation: WKB approximation}
\label{sec:analytics}

We solve Equation~(\ref{eq:kink}) analytically by using the Wentzel-Kramers-Brillouin (WKB) approximation \citep[see, e.g.,][for details about the method]{bender}. The WKB approximation has been recently applied to the investigation of MHD waves in cooling coronal loops \citep{morton1,morton2,morton3}. In particular, the work by \citet{morton2} is especially relevant for the present investigation as they studied kink oscillations of coronal loops with variable background.  

The WKB approximation is an approximate method to study waves in a changing background whose properties are smooth functions of space and/or time. In the present application of the WKB approximation we assume that the time scale related to the waves, e.g., the period, is much shorter than the time scale related to the changes of the background configuration. Under these conditions, it is possible to define a {\em time-dependent frequency} which slowly varies because of the changing background. To apply the WKB approximation we define the parameter $\delta$ as
\begin{equation}
 \delta \equiv \frac{v_0}{L}. \label{eq:deltadef}
\end{equation}
The validity of the WKB approximation is restricted to small values of  $\delta$ so as $P \delta  \ll 1$, where $P$ is the period of the oscillations. In the observations by \citet{okamoto}, the mean flow velocity and period are $v_0 \approx 40$~km~s$^{-1}$ and $P\approx 3$~min. For $L \sim 10^5$~km these values result $P \delta  \approx 0.072$, meaning that the condition of applicability of the WKB approximation is fulfilled in the case of transverse oscillations of flowing threads.

Using the parameter $\delta$ we define the  dimensionless time, $t_1$, as
\begin{equation}
 t_1= \delta t,
\end{equation}
and we express the solution to Equation~(\ref{eq:kink}) in the following form
\begin{equation}
v_r(z,t_1) = Q_1 (z,t_1) \exp \left( \frac{i}{\delta} \Omega_1 (t_1) \right), \label{eq:wkb}
\end{equation}
with $Q_1 (z,t_1)$ and $\Omega_1 (t_1)$ functions to be determined. Next, we combine Equations~(\ref{eq:kink}) and (\ref{eq:wkb}), and separate the different terms according to their order with respect to $\delta$. As $\delta$ is small, the dominant terms are those with the lowest order in $\delta$. We obtain two equations for $Q_1 (z,t_1)$ and $\Omega_1 (t_1)$ taking the terms with $\mathcal{O} \left( \delta^0 \right)$ and $\mathcal{O}\left( \delta^1 \right)$, namely
\begin{eqnarray}
 \frac{\pd^2 Q_1 (z,t_1) }{\pd z^2} + \left( \frac{\pd \Omega_1 (t_1) }{\pd t_1} \right)^2 \frac{Q_1 (z,t_1)}{\ck^2(z,t_1)} &=& 0, \label{eq:basic1} \\
Q_1 (z,t_1)\frac{ \pd^2 \Omega_1 (t_1) }{\pd t_1^2} + 2 \frac{\pd Q_1 (z,t_1) }{\pd t_1} \frac{\pd \Omega_1 (t_1) }{\pd t_1} &=& 0.\label{eq:basic2}
\end{eqnarray}
Equations~(\ref{eq:basic1}) and (\ref{eq:basic2}) are equivalent to Equations~(24) and (25) of \citet{morton2}, respectively.

Now, we define the time-dependent frequency, $\omega \left( t_1 \right)$, as
\begin{equation}
 \omega \left( t_1 \right) \equiv  \frac{\pd \Omega_1 (t_1) }{\pd t_1}, \label{eq:frequencydef}
\end{equation}
which allows us to rewrite Equation~(\ref{eq:basic1}) as follows
\begin{equation}
 \frac{\pd^2 Q_1 (z,t_1) }{\pd z^2} + \frac{\omega^2 \left( t_1 \right)}{\ck^2(z,t_1)} Q_1 (z,t_1) = 0. \label{eq:kinkQ}
\end{equation}
 Equation~(\ref{eq:kinkQ}) has to be solved taking into account the boundary conditions $Q_1 ( \pm L/ 2 ,t_1) = 0$ due to photospheric line-tying. By solving Equation~(\ref{eq:kinkQ}), the dependence on $z$ of function $Q_1 (z,t_1)$ can be obtained. In addition, since $\ck^2(z,t_1)$ is a piecewise constant function of $z$ (see Equation~(\ref{eq:kinkspeed})), the analytical solutions to Equation~(\ref{eq:kinkQ}) are trigonometric functions with time-dependent arguments. Thus, the general solution to Equation~(\ref{eq:kinkQ}) satisfying the boundary conditions at $z = \pm L/2$ is
\begin{equation}
 Q_1  (z,t_1)  =  \left\{
\begin{array}{lll}
 A_1( t_1 ) \sin \left( \frac{\omega \left( t_1 \right)}{\cke} \left( z + \frac{L}{2} \right) \right) & \textrm{if} & z < z_-, \\
A_2(t_1) \cos \left( \frac{\omega \left( t_1 \right)}{\ckp}  z + \phi(t_1)  \right) & \textrm{if} & z_-\leq z  \leq z_+, \\
 A_3( t_1 ) \sin \left( \frac{\omega \left( t_1 \right)}{\cke} \left( z - \frac{L}{2} \right) \right) & \textrm{if} & z  > z_+ 
\end{array} \right. \label{eq:solgen}
\end{equation}
where $A_1( t_1 )$, $A_2(t_1)$, and $A_3(t_1)$ are time-dependent coefficients, $\phi( t_1 )$ is a time-dependent phase, and $z_-$ and $z_+$ denote the locations of the interfaces between the prominence thread and the evacuated regions, namely
\begin{equation}
z_{\pm} = z_0 \pm \frac{\lp}{2} + t_1 L.
\end{equation}
The locations of the interfaces change as the dense thread moves along the magentic tube. $Q_1$ must satisfy appropriate boundary conditions at $z=z_\pm$. Since the interfaces correspond to contact discontinuities \citep[see][]{goedbloed}, the boundary conditions are
\begin{equation}
\left[\left[Q_1  \right]\right] = 0, \quad  \left[\left[ \frac{\pd Q_1  }{\pd z}  \right]\right] = 0,\label{eq:boundary}
\end{equation}
where $[[X]]$ stands for the jump of the quantity $X$ at $z = z_{\pm}$.

Applying the conditions of Equation~(\ref{eq:boundary}) on the solutions given by Equation~(\ref{eq:solgen}), we arrive at the following equation
\begin{eqnarray} 
&& \frac{\cke}{\ckp} \tan \left[ \frac{\omega\left( t_1 \right)}{\cke} \left( z_0 - \frac{L-\lp}{2} + t_1 L  \right) \right] \nonumber \\
&=& \frac{\ckp + \cke \cot \left( \frac{\omega\left( t_1 \right)}{\ckp} \lp \right) \tan \left[ \frac{\omega\left( t_1 \right)}{\cke} \left(z_0 + \frac{L-\lp}{2} + t_1 L \right) \right] }{\ckp \cot \left( \frac{\omega\left( t_1 \right)}{\ckp} \lp \right) - \cke  \tan \left[ \frac{\omega\left( t_1 \right)}{\cke} \left(z_0 + \frac{L-\lp}{2} + t_1 L \right) \right]}.
\label{eq:disper}
\end{eqnarray}
Equation~(\ref{eq:disper}) is the time-dependent dispersion relation. For fixed $t_1$, the solution of Equation~(\ref{eq:disper}) is $\omega \left( t_1 \right)$. Note that, although Equation~(\ref{eq:disper}) is written in a more compact form, it is consistent with dispersion relations previously obtained for normal modes in the static case, i.e., $v_0 = 0$. Equation~(\ref{eq:disper}) with $v_0 = 0$ is equivalent to Equation~(11) of \citet{solerstatic} if the substitutions $L_{\rm e}^+ \to \frac{L-\lp}{2} - z_0$ and $L_{\rm e}^- \to \frac{L-\lp}{2} + z_0$ are performed in their expression. Also for $v_0 = 0$, Equation~(\ref{eq:disper}) is similar to Equation~(17) of \citet{joarder} and Equation~(A5) of \citet{oliverslab} obtained in Cartesian geometry.

\begin{figure}[!t]
\centering
 \includegraphics[width=0.99\columnwidth]{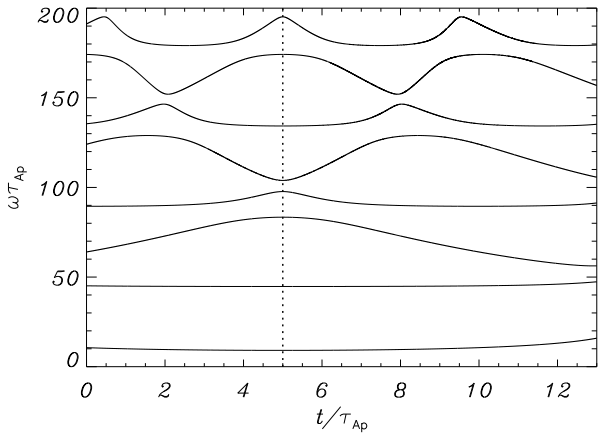}
\caption{Dimensionless frequency, $\omega \ta$, versus time in units of the internal Alfv\'en travel time, $\ta = L / \vap$. Results corresponding to the fundamental mode and the lowest seven harmonics obtained by numerically solving Equation~(\ref{eq:disper}) for a flowing thread with $\lp / L =$~0.1, $v_0 / \vap =$~0.05, and $z_0 / L = -0.25$. The vertical dotted line denotes the time when the prominence thread is centered within the flux tube. \label{fig:disper}}
\end{figure}

We have solved Equation~(\ref{eq:disper}) by standard numerical techniques. The frequencies of the fundamental mode and of the lowest seven harmonics with respect to $z$ are displayed as functions of time  in Figure~\ref{fig:disper} for a particular set of parameters. We find that the dispersion diagram is symmetric with the time when the thread is located at the center of the magnetic tube (denoted by a vertical dotted line in Fig.~\ref{fig:disper}) as point of symmetry. The fundamental mode and the first harmonic are smooth functions of time. The other harmonics displayed in Figure~\ref{fig:disper} show a complicated set of couplings and avoided crossings. The reason for this behavior is that the fundamental mode and the first harmonic correspond to {\em global} oscillations of the flux tube because both the prominence and the evacuated parts of the tube are disturbed. On the contrary, high harmonics correspond to modes more confined within one of these regions. Thus, the collection of modes and their properties are similar to those studied by \citet{joarder} and \citet{oliverslab} in slab geometry.

\subsection{Approximation to the fundamental mode frequency}

From hereon we restrict our analysis to the fundamental mode of oscillation, whose frequency is the lowest order solution to Equation~(\ref{eq:disper}). To obtain an approximation to the frequency, we perform a Taylor expansion of Equation~(\ref{eq:disper}) and neglect terms with $\mathcal{O} \left( \omega^4 \right)$ and higher orders in $\omega$. The following expression is obtained
\begin{eqnarray}
& &\omega \left( t_1 \right) \approx 2 \ckp \sqrt{ \frac{L}{\lp} }  \nonumber \\  &\times & \frac{1}{\sqrt{ \left( L - \lp \right) \left( L + \frac{1}{3} \lp \right) - 4 \left( z_0 + t_1 L \right)^2 \left( 1 + \frac{\re+\rc}{\rp+\rc} \frac{\left( L - \lp \right)}{\lp} \right) } }. \label{eq:freq}
\end{eqnarray}

The effect of the flow is contained in the denominator of the right-hand side of Equation~(\ref{eq:freq}).  We see that the effect of the flow on the frequency is more complicated than a simple Doppler shift. There are two reasons that cause this dependence. On the one hand, our model is a complicated structure in the sense that only the dense prominence material is moving. It is well known that a wave propagating in a uniform magnetic tube with a constant siphon flow is affected by a constant Doppler shift of the frequency due to the flow. However, the effect of the flow is not so simple in more complicated configurations. Even in the case of a flux tube with a constant flow within the tube but no flow in the exterior of the tube the wave frequencies suffer corrections due to the flow that are not simple frequency shifts \citep[see, e.g.,][]{nakaroberts,terra}. Our configuration is very different from the typical uniform magnetic flux tube with a siphon flow, so that the frequency is also modified by the change of position of the dense plasma within the magnetic tube. On the other hand, we are dealing with standing modes, not propagating waves. For standing modes in flux tubes \citet{terradasflow} have shown that flow produces a spatially dependent phase shift along the magnetic tube. In our case, this phase shift is contained in the time-dependent phase $\phi( t_1 )$ of Equation~(\ref{eq:solgen}).

For typical prominence and coronal densities, $\rp \gg \rc$ and $\rp \gg \re$. Therefore, the term with the ratio of densities in the denominator of Equation~(\ref{eq:freq}) can be neglected. Then, Equation~(\ref{eq:freq}) simplifies to
\begin{equation}
 \omega \left( t_1 \right) \approx   \frac{2 \ckp \sqrt{\frac{L}{\lp}}}{\sqrt{ \left( L - \lp \right) \left( L + \frac{1}{3} \lp \right) - 4 \left( z_0 + t_1 L \right)^2  } }. \label{eq:freq2} 
\end{equation}
In the absence of flow, i.e., $t_1 = 0$, and for $z_0 =0$, Equation~(\ref{eq:freq2}) loses its time dependence and becomes,
\begin{equation}
 \omega  \approx  2 \ckp \sqrt{ \frac{L}{ \left( L - \lp \right) \left( L + \frac{1}{3} \lp \right) \lp } }. \label{eq:freq3} 
\end{equation}
Equation~(\ref{eq:freq3}) is consistent with the approximation of the normal mode frequency obtained by \citet[Equation~(8a)]{diazperiods}. For $\lp \ll L$ we can approximate $L + \frac{1}{3} \lp \approx L$, and Equation~(\ref{eq:freq3}) reduces to the expression found by \citet[Equation~(17)]{solerstatic}. Although the  case $\lp \to L$ is very unrealistic in prominences because the observed lengths of prominence threads correspond to $\lp / L \ll 1$, it is instructive to take into account this limit.  For $\lp \to L$ the frequency given in Equation~(\ref{eq:freq3}) tends to infinity. The reason is that the fundamental kink mode behaves as a {\em hybrid} mode like those described by \citet{oliverslab} in a Cartesian slab \citep[see also the {\em string modes} investigated by][]{joarder}. Equations~(\ref{eq:freq})--(\ref{eq:freq3}) are approximations of the {\em hybrid} mode frequency. As explained by \citet{oliverslab}, {\em hybrid} modes owe their existence to the presence of both the dense part and the evacuated part of the tube. In the limit $\lp \to L$ the evacuated part is absent and the {\em hybrid} kink mode disappears. Thus for $\lp \to L$ the fundamental kink mode is not the {\em hybrid} mode but the first {\em internal} mode with frequency
\begin{equation}
 \omega = \ckp \frac{\pi}{L}.
\end{equation}
Since $\lp / L \ll 1$ in prominences, the fundamental mode in a realistic thread is the {\em hybrid} mode and Equations~(\ref{eq:freq})--(\ref{eq:freq3}) apply.

\begin{figure*}[!t]
\centering
 \includegraphics[width=0.99\columnwidth]{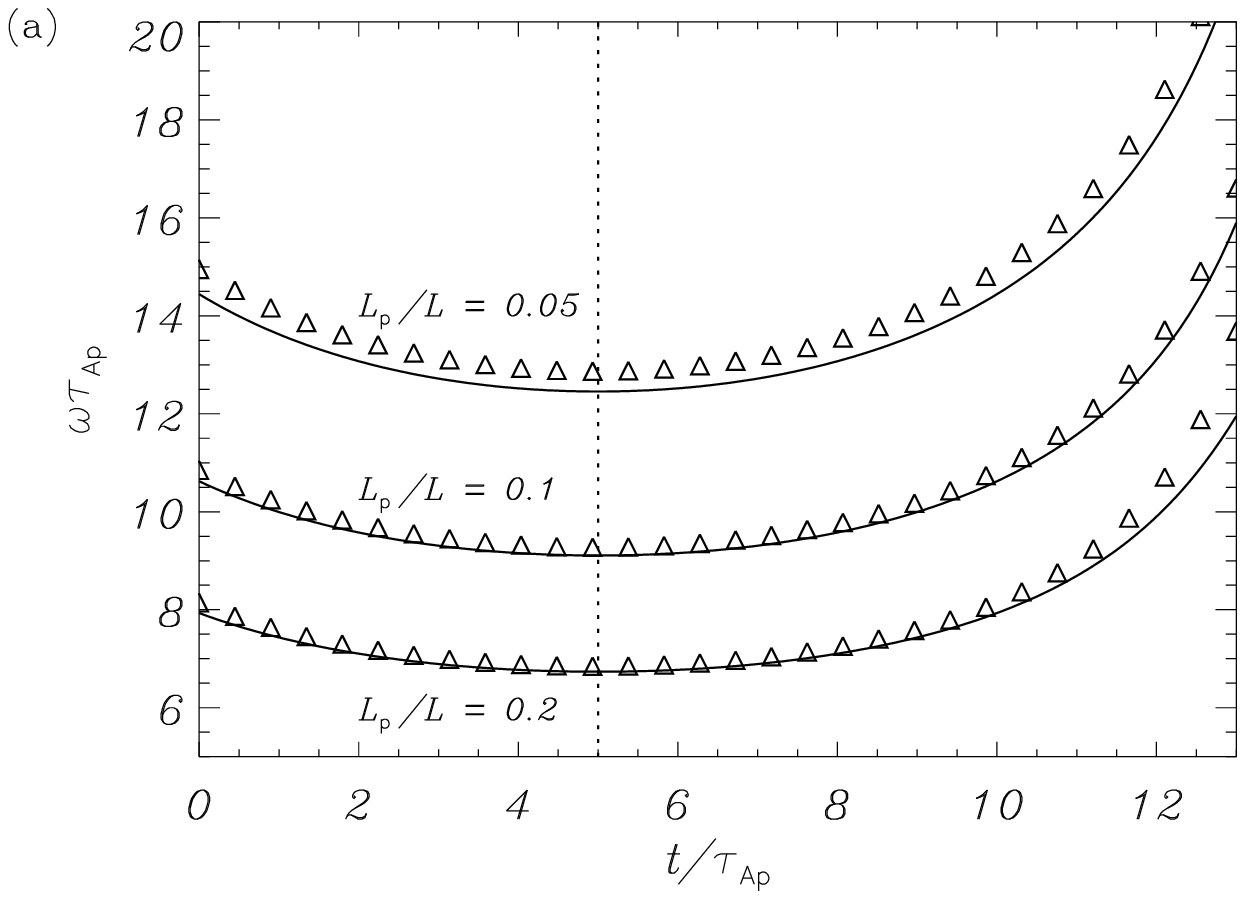}
  \includegraphics[width=0.99\columnwidth]{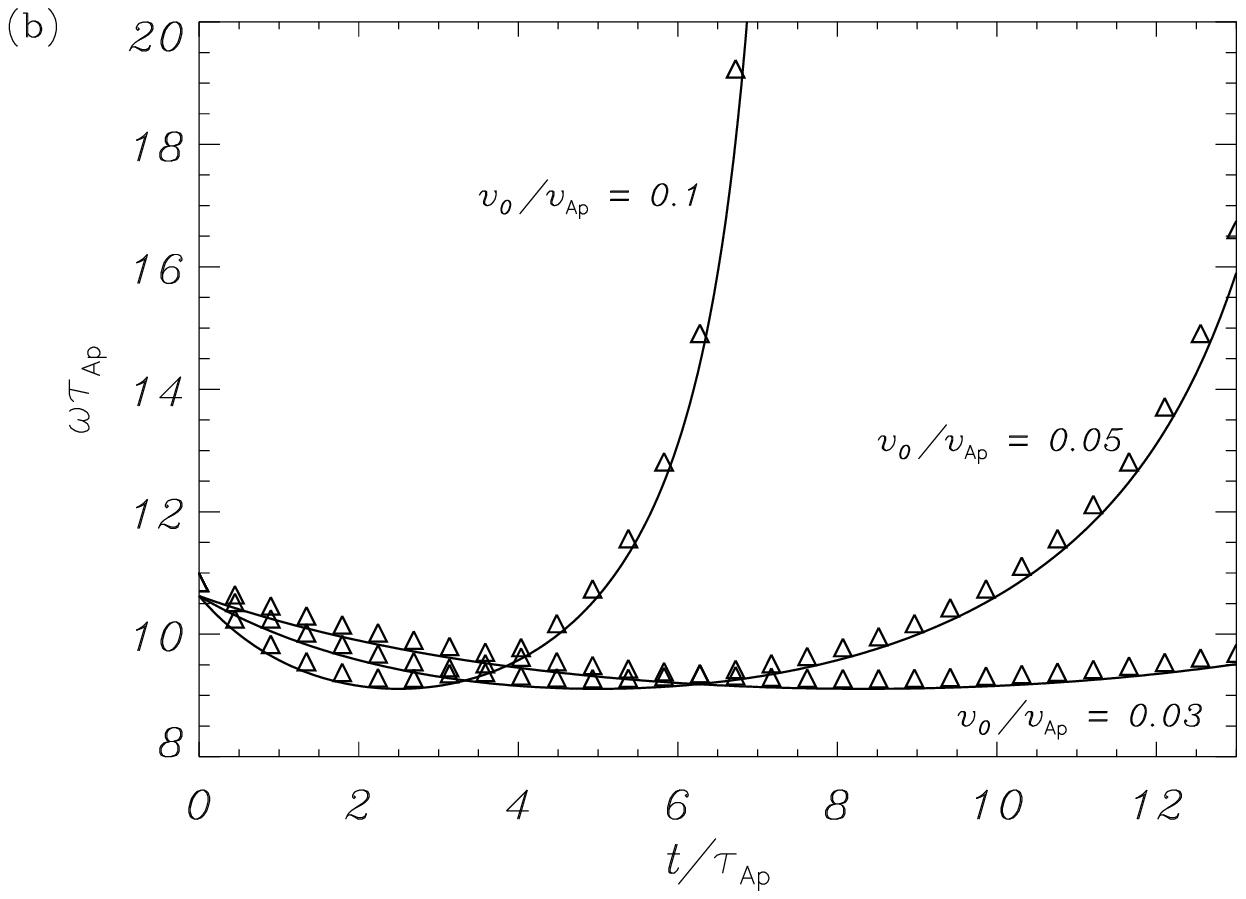}
\caption{Dimensionless frequency, $\omega \ta$, versus time in units of the internal Alfv\'en travel time, $\ta = L / \vap$, for (a) $\lp / L =$~0.05, 0.1, 0.2 with $v_0 / \vap =$~0.05, and (b) $v_0 / \vap =$~0.03, 0.05, 0.1 with $\lp / L = 0.1$. In all cases, $z_0 / L = -0.25$. The solid lines are the results obtained by numerically solving Equation~(\ref{eq:disper}), whereas the symbols correspond to the analytical approximation given by Equation~(\ref{eq:freq2}). The vertical dotted line in panel (a) denotes the minimum of the curves, i.e., the time when the prominence thread is centered within the flux tube. \label{fig:comp}}
\end{figure*}

We plot in Figure~\ref{fig:comp} the solution of Equation~(\ref{eq:disper}) for the fundamental mode as a function of time, along with the analytical approximation given by Equation~(\ref{eq:freq2}). Equation~(\ref{eq:disper}) has been solved by using standard numerical methods to obtain the roots of transcendental equations. These computations have been performed for different values of $\lp / L$ and $v_0 / \vap$, and for fixed $z_0 / L$. There is a very good agreement between the full solution (solid lines in Figure~\ref{fig:comp}) and the approximation (symbols). From Figure~\ref{fig:comp}(a), we see that the frequency decreases as the length of the prominence thread increases. On the other hand, Figure~\ref{fig:comp}(b) shows that the variation of the frequency with time is more important as the flow velocity gets faster. We find that the minimum of the frequency takes place when the thread is centered within the magnetic tube. Therefore, the minimum of the frequency depends on both the initial position of the thread and the flow velocity since the relation $z_0 + v_0 t = 0$ has to be satisfied.

Equation~(\ref{eq:freq2}) corresponds to the instantaneous frequency. However, the actual temporal dependence of the oscillation in the WKB approximation is given by function $\Omega_1(t_1)$. We obtain $\Omega_1(t_1)$ from Equation~(\ref{eq:frequencydef}) by integrating $\omega \left( t_1 \right)$ given by Equation~(\ref{eq:freq2}). Hence,
\begin{eqnarray}
 \Omega_1(t_1) &=& \frac{\ckp}{\sqrt{L \lp}}\left\{ \arctan \left[ \frac{2 \left( z_0 + t_1 L  \right)}{\sqrt{\left( L - \lp  \right) \left( L + \frac{1}{3} \lp \right) - 4 \left( z_0 + t_1 L  \right)^2}}  \right] \right.  \nonumber \\ &-& \left. \arctan \left[ \frac{2  z_0 }{\sqrt{\left( L - \lp  \right) \left( L + \frac{1}{3} \lp \right) - 4 z_0^2}}  \right] \right\},
\end{eqnarray}
where we have used the condition $\Omega_1 = 0$ at $t_1 = 0$.

\subsection{Dependence of the amplitude on time}
\label{sec:amplitude}

Here, we estimate the variation of the amplitude of the oscillations with time. To do so, we use Equation~(\ref{eq:basic2}). By taking into account the definition of the time-dependent frequency (Equation~(\ref{eq:frequencydef})), we rewrite Equation~(\ref{eq:basic2}) as
\begin{equation}
 \frac{\pd Q_1 (z,t_1) }{\pd t_1} + \frac{1}{2 \omega (t_1)} \frac{ \pd \omega (t_1) }{\pd t_1} Q_1 (z,t_1) = 0.
\end{equation}
Next, we consider the approximate $\omega (t_1)$  for the fundamental mode obtained in Equation~(\ref{eq:freq}) to express this last Equation as
\begin{eqnarray}
 \frac{\pd Q_1 (z,t_1) }{\pd t_1} + \frac{2 L \left( z_0 + t_1 L \right)}{\left( L - \lp \right) \left( L + \frac{1}{3} \lp \right) - 4 \left( z_0 + t_1 L \right)^2}Q_1 (z,t_1) = 0. \nonumber \\ \label{eq:qtempdep}
\end{eqnarray}
Note that to solve Equation~(\ref{eq:qtempdep}) we do not have to care about the dependence of $Q_1$ on $z$. For a given $z$, Equation~(\ref{eq:qtempdep}) can be integrated to obtain the temporal dependence of $Q_1$ at a fixed position, namely
\begin{eqnarray}
 Q_1 (z,t_1) &=& Q_0(z) \left[ \frac{\left( L - \lp \right) \left( L + \frac{1}{3} \lp \right) - 4 \left( z_0 + t_1 L \right)^2}{\left( L - \lp \right) \left( L + \frac{1}{3} \lp \right) - 4 z_0^2} \right]^{1/4},\label{eq:qtempdepsol}
\end{eqnarray}
where $Q_0(z)$ is the amplitude at $t_1=0$. Thus, Equation~(\ref{eq:solgen}) gives the spatial dependence of $Q_1$ for a fixed time $t_1$, while Equation~(\ref{eq:qtempdepsol}) provides the temporal dependence at a fixed position $z$. By comparing Equations~(\ref{eq:freq2}) and (\ref{eq:qtempdepsol}), we see that $Q_1 (z,t) \propto \omega\left( t_1 \right)^{-1/2}$. Then, the amplitude of the oscillation decreases when the thread flows from the center of the tube to the footpoint, and increases otherwise. This means that the amplitude is maximal when the thread is located at the center of the magnetic tube.  

We must bear in mind that Equation~(\ref{eq:qtempdepsol}) was derived from the terms with $\mathcal{O}\left( \delta^1 \right)$ in the governing equation, whereas the leading terms when $\delta$ is a small parameter are those with $\mathcal{O}\left( \delta^0 \right)$. Therefore, we have to be cautious about the actual accuracy of Equation~(\ref{eq:qtempdepsol}), although we expect the behavior of the amplitude with time to be, at least, qualitatively described by Equation~(\ref{eq:qtempdepsol}).

\subsection{Application to magneto-seismology}

The analytical expression of the fundamental mode frequency found before can be used to perform magneto-seismology of prominence fine structures by using observed periods of oscillations in flowing threads. We use Equation~(\ref{eq:freq2}) to compute the period of the oscillation as a function of time, namely
\begin{eqnarray}
 P_1 \left( t \right) &=& \frac{2 \pi}{\omega \left( t \right)} \nonumber \\
&\approx& \frac{\pi}{\ckp} \sqrt{\frac{\lp}{L}} \sqrt{\left( L - \lp \right) \left( L + \frac{1}{3} \lp \right) - 4 \left( z_0 + v_0 t \right)^2}, \label{eq:period}
\end{eqnarray}
which we have explicitly written in terms of the dimensional time, $t$, and the flow velocity, $v_0$. By using Equation~(\ref{eq:period}) along with observational values of the period, it is possible to give an estimation of $L$, i.e., the total length of the flux tube, which is a parameter difficult to measure from the observations. Let us assume that we have performed an observation of a transversely oscillating and flowing thread with a good cadence and we have determined the evolution of the period with time. For convenience, we set $t=0$ when the maximum of the period takes place, namely $P_1 (0)$, so we can also fix $z_0=0$ in Equation~(\ref{eq:period}). Then, we denote as $P_1 (\tau)$ the instantaneous period measured at $t=\tau$. We use Equation~(\ref{eq:period}) and compute the ratio $P_1 (\tau) / P_1 (0)$ to find an estimation of the length of the magnetic flux tube as
\begin{equation}
 L \approx \lp + \frac{v_0^2 \tau^2}{\lp} \frac{2}{1 - \left( \frac{P_1 (\tau)}{P_1 (0)} \right)^2}, \label{eq:length}
\end{equation}
where again we assumed that the flow velocity is slow. We see that the right-hand side of Equation~(\ref{eq:length}) depends on quantities that can be directly measured from the observations. As an example, let us assume the following values: $P_1 (\tau) / P_1 (0) = 0.9$, $v_0 = 40$~km~s$^{-1}$, $\lp = 10,000$~km, and $\tau = 5$~min. From Equation~(\ref{eq:length}) we obtain $L \approx 1.6 \times 10^5$~km. However, the accuracy of Equation~(\ref{eq:length}) is limited by the uncertainties and error bars of the observations. In particular, a very accurate determination of the ratio $P_1 (\tau) / P_1 (0)$ is needed. For instance, a 10\% uncertainty of $P_1 (\tau) / P_1 (0)$ produces a 85\% uncertainty of $L$ when propagation of errors is used in Equation~(\ref{eq:length}) and the remaining parameters are kept constant. This makes a reliable application of Equation~(\ref{eq:length}) difficult in practice.

\begin{figure}[!t]
\centering
 \includegraphics[width=0.99\columnwidth]{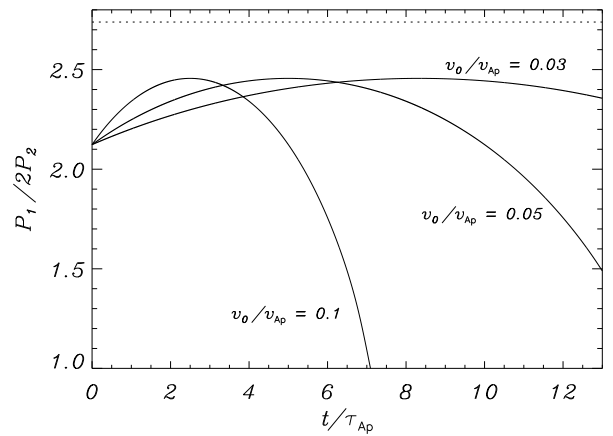}
\caption{Ratio $P_1 / 2 P_2$ of a flowing thread with $\lp / L = 0.1$ and $z_0 / L=-0.25$ for $v_0 / \vap =$~0.03, 0.05, 0.1. The dotted line is the result from Equation~(\ref{eq:periodsdiaz}) in the case without flow and for a prominence thread located at the center of the magnetic structure. \label{fig:periodrat}}
\end{figure}

Another relevant parameter that can give us a seismological determination of $L$ is the ratio $P_1 / 2 P_2$, where $P_2$ is the period of the first harmonic. The deviation of this ratio from unity is an indication of the longitudinal inhomogeneity length scale of the magnetic tube. Its application was used for the first time in the context of coronal loop oscillations by \citet{andries2005a,andries2005b} and has been explored in subsequent works \citep[see the recent review by][and references therein]{andriesrev}. In prominence thread oscillations, \citet{diazperiods} explored the importance of the ratio $P_1 / 2 P_2$ to estimate $\lp / L$ in static threads. While in coronal loops $P_1 / 2 P_2 < 1$, \citet{diazperiods} found that in prominence threads $P_1 / 2 P_2 > 1$. The reason for this result is that in prominence threads mass density is arranged just in the opposite way to that in coronal loops. In loops the density is larger in the footpoints than in the apex due to gravitational stratification, while in prominence threads the density is much larger in the center of the magnetic tube because of the presence of the prominence material.

For a large density contrast between the prominence and the coronal plasmas, and assuming that the thread is located at the center of the magnetic cylinder, the relation between $P_1 / 2 P_2$ and $\lp / L$ obtained by \citet{diazperiods} in their Equation~(11) is
\begin{equation}
 \frac{P_1}{2 P_2} \approx \sqrt{\frac{3}{4 \lp / L}}. \label{eq:periodsdiaz}
\end{equation}
Let us see how the ratio $P_1 / 2 P_2$ is affected by the flow and so how the results of \citet{diazperiods} are modified. However, it is difficult to obtain an analytical expression for $P_2$ when flow is present. Instead, we compute both $P_1$ and $P_2$ by solving Equation~(\ref{eq:disper}) with numerical methods. Figure~\ref{fig:periodrat} shows $P_1 / 2 P_2$ as a function of time for $\lp / L = 0.1$ and $z_0 / L=-0.25$, and for different values of $v_0 / \vap$. The numerical results are compared with the analytical expression of \citet{diazperiods}. First of all, we see that the period ratio is strongly influenced by the velocity at which the prominence thread flows along the flux tube.  $P_1 / 2 P_2$ is maximal when the thread in centered within the tube ($P_1 / 2 P_2 \approx 2.5$ for the particular set of parameter in Fig.~\ref{fig:periodrat}) and $P_1 / 2 P_2 \to 1$ when the thread reaches the footpoint. The analytical approximation of \citet{diazperiods} for the static case gives a larger value of $P_1 / 2 P_2$ in comparison to the case with flow. This means that flow reduces the period ratio. Therefore, Equations~(\ref{eq:length}) and (\ref{eq:periodsdiaz}) could be used together to obtain a more accurate determination of the magnetic tube length, as the value of $L$ inferred from Equation~(\ref{eq:periodsdiaz}) should be considered as a upper bound for this parameter.

\section{Numerical results: time-dependent simulations}
\label{sec:numerics}

Here, we compare the analytical results of the WKB approximation with the full numerical solution of the time-dependent problem. We use the PDE2D code \citep{sewell} for that purpose. The set-up of the numerical code is similar to that of \citet{hinode}. Equation~(\ref{eq:kink}) is integrated assuming the boundary conditions $v_r(\pm L / 2,t) = 0$. An initial condition for $v_r$ at $t=0$ is also provided. In the code, the distances are expressed in units of $L$ and the velocities in units of the prominence Alfv\'en speed, $\vap = \frac{B}{\sqrt{\mu \rp}}$. We take $L=10^5$~km. Assuming $B=50$~G and $\rp=10^{-10}$~kg~m$^{-3}$ as realistic values in active region prominences, we obtain $\vap \approx 446$~km~s$^{-1}$. The flow velocities on the plane of sky estimated by \citet{okamoto} are in the interval between 15~km~$^{-1}$ to 46~km~s$^{-1}$. Hence in our simulations we consider values for the ratio $v_0 / \vap$ in the range $0.03 \lesssim v_0 / \vap \lesssim 0.1$. In the code, time is expressed in units of the Alfv\'en travel time, i.e., $\ta = L / \vap \approx 3.74$~min. In addition, in all the following computations we have used $\rp / \rc = 200$ and $\re / \rc = 1$.

\begin{figure*}[!t]
\centering
 \includegraphics[width=0.79\columnwidth]{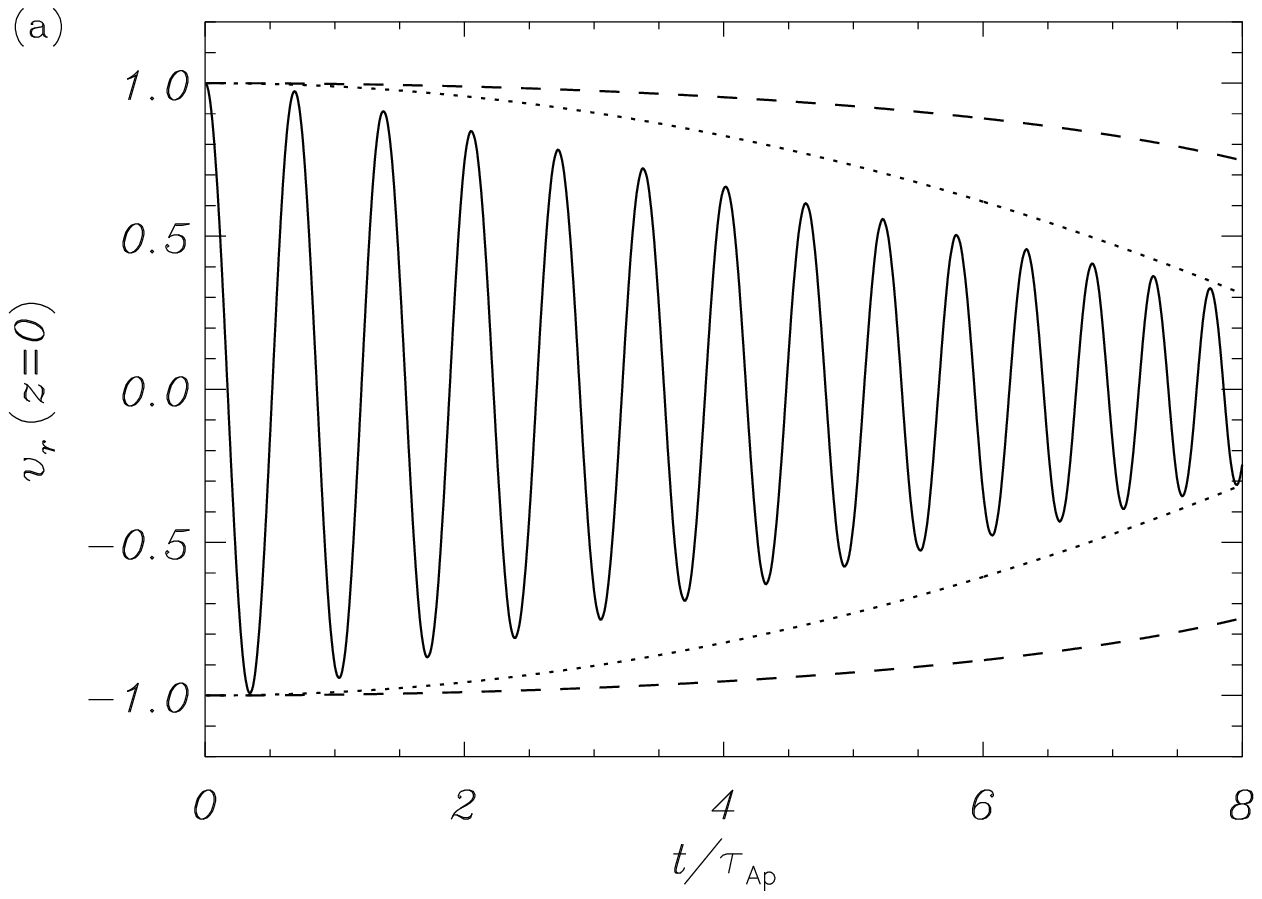}
 \includegraphics[width=0.79\columnwidth]{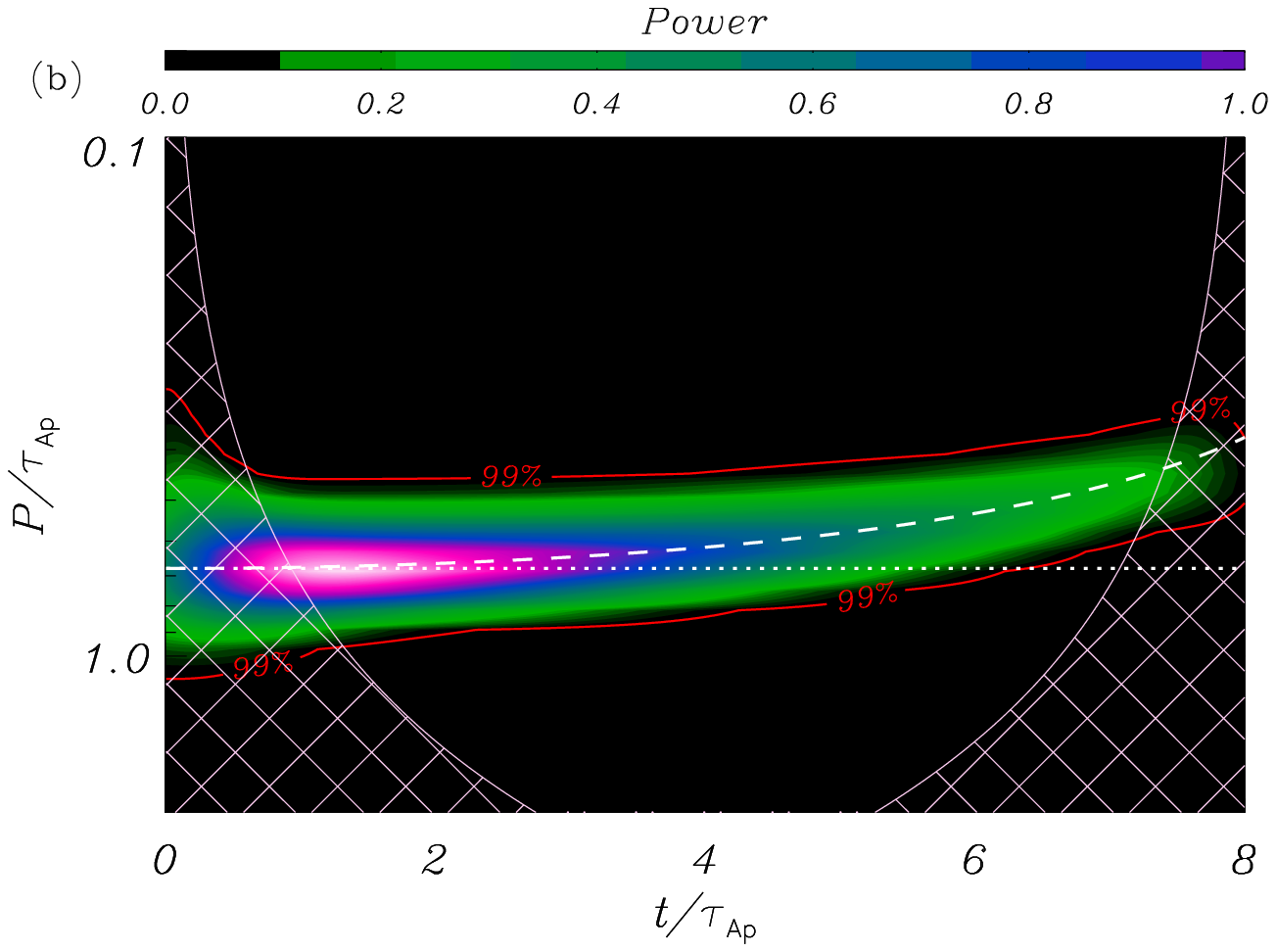}
\includegraphics[width=0.79\columnwidth]{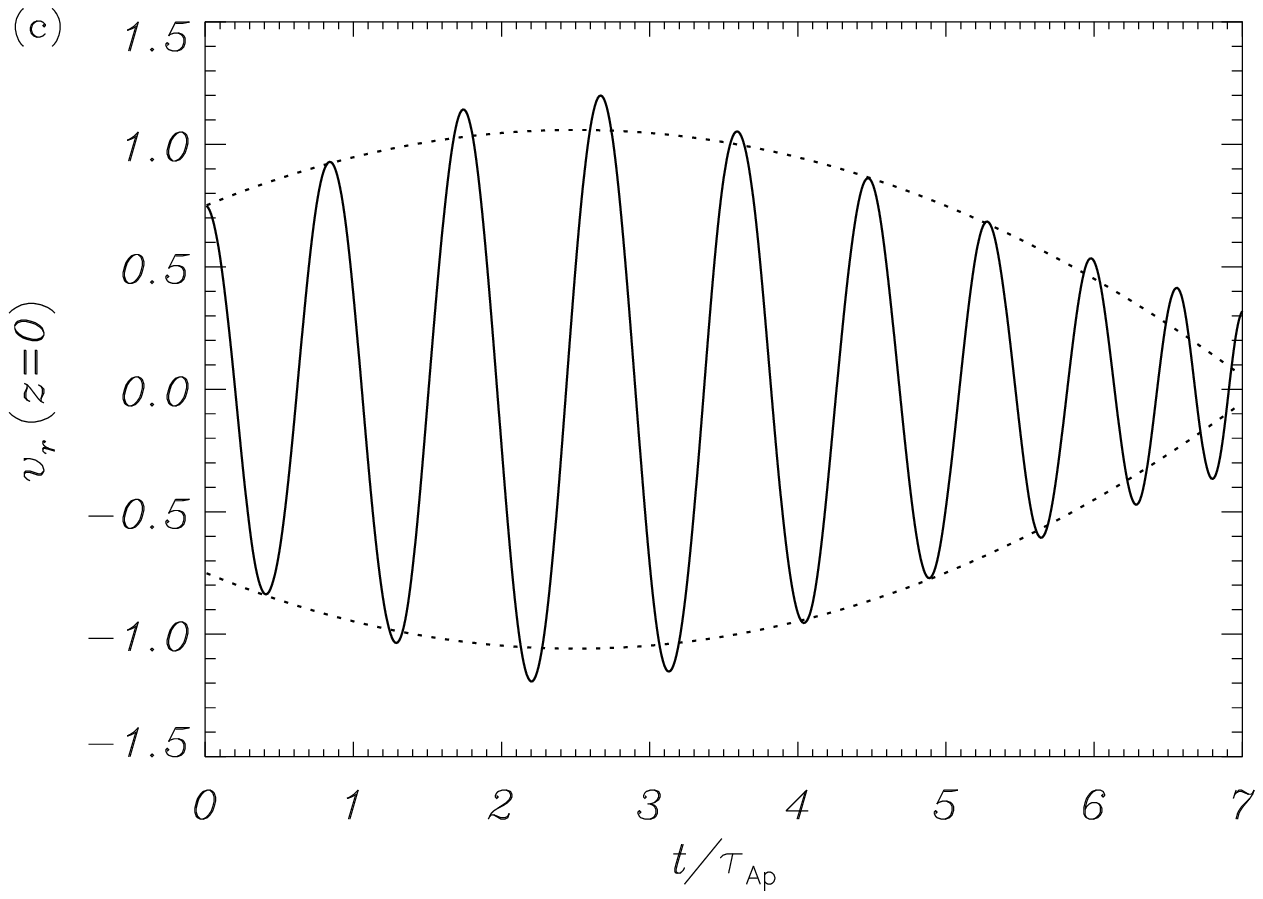}
  \includegraphics[width=0.79\columnwidth]{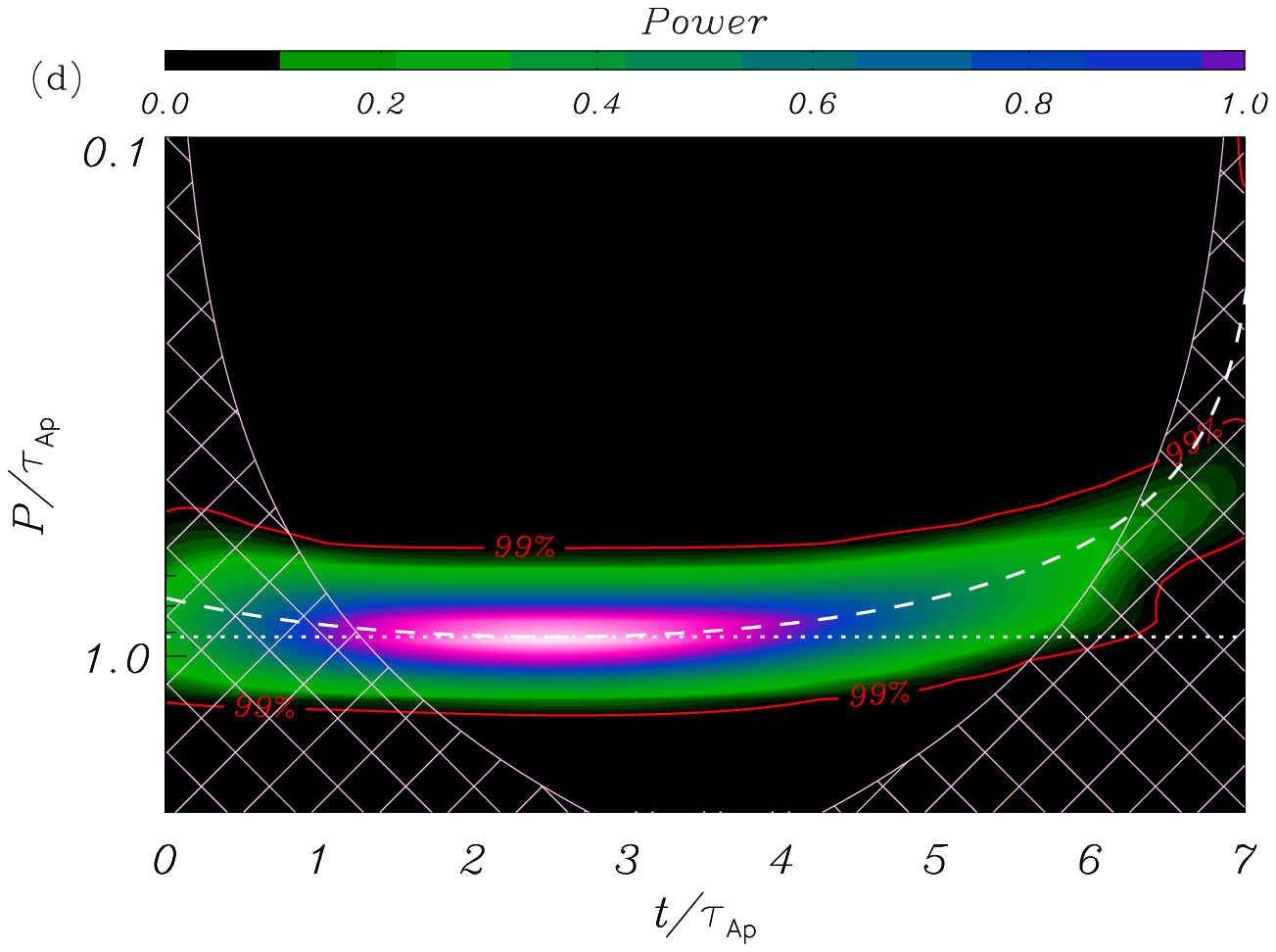}
\caption{(a) $v_r$ at $z=0$ (solid line in arbitrary units) as a function of time in a prominence fine structure with $\lp / L = 0.1$, $v_0 / \vap = 0.05$, and $z_0=0$. The initial excitation is the fundamental kink mode eigenfunction for $v_0 = 0$. The dashed line is the amplitude in the WKB approximation (Equation~(\ref{eq:qtempdepsol})), while the dotted line corresponds to the fit proposed in Equation~(\ref{eq:fit}) with $n=1$. (b) Wavelet power spectrum for the dimensionless period, $P / \ta$, corresponding to the signal displayed in panel (a). The white dashed line is the period in the WKB approximation (Equation~(\ref{eq:period})), whereas the horizontal dotted line is the period at $t=0$. The red solid line denotes 99\% of confidence level. (c) Same as panel (a) but for $\lp / L = 0.2$, $v_0 / \vap = 0.1$, and $z_0 / L = -0.25$. (d) Same as panel (b) but for the signal of panel (c), and with the horizontal dotted line denoting the maximum of the period. \label{fig:flow1}}
\end{figure*}

\subsection{Excitation of the fundamental mode}

First, we use the eigenfunction of the fundamental kink mode as the initial condition for $v_r$ at $t=0$. The eigenfunction is obtained by solving the dispersion relation of the normal mode problem and computing the spatial distribution of the corresponding perturbation \citep[see details in][]{dymovaruderman, solerstatic, diazperiods}. Hence, we make sure that, after the initial excitation, the magnetic tube mainly oscillates in its fundamental mode. 

As a check of the numerical code, we consider the static case and put the thread at the center of the magnetic tube, i.e., $v_0 = 0$ and $z_0=0$. In this test simulation, we take $\lp / L = 0.1$. By looking at the time-dependent evolution of $v_r$, we check that the magnetic tube oscillates as a whole. A plot of $v_r$ at $z=0$ as a function of time (not displayed here for the sake of simplicity) shows that the amplitude of the oscillation is constant during the whole duration of the simulation, meaning that numerical dissipation is negligible in the simulation. Later, we perform a power spectrum of $v_r$ at $z=0$ \citep[see, e.g.,][]{carball} and find a large peak centered around the fundamental normal mode frequency. The maximum of the power spectrum agrees very well with the approximate frequency of the normal mode given by Equation~(\ref{eq:freq2}). For the set of parameters used in this numerical test, the period in dimensional units is $P \approx 2.6$~min.  We have performed similar simulations but using different values of $\lp / L$ and $z_0$. Equivalent results to those commented before have been obtained in all cases. This indicates, on the one hand, that the normal mode interpretation is a very good representation for the time-dependent evolution of the oscillation in the static case and, on the other hand, that the numerical code works properly.

Hereafter, we incorporate the effect of the flow. First, we fix $\lp / L = 0.1$ and $z_0 = 0$, and consider $v_0 / \vap = 0.05$. In this situation, the thread is initially located at the center of the magnetic tube. We make sure that the simulation stops before the threads reaches the photospheric wall. We plot in  Figure~\ref{fig:flow1}(a) the radial velocity perturbation at $z=0$ as a function of time. Contrary to the static case, now the amplitude of the oscillation decreases in time as the thread moves from the center towards the end of the magnetic tube. This behavior is qualitatively described by Equation~(\ref{eq:qtempdepsol}) obtained in the WKB approximation (see dashed line in Fig.~\ref{fig:flow1}(a)), although Equation~(\ref{eq:qtempdepsol}) underestimates the actual decrease of the amplitude. Inspired by   Equation~(\ref{eq:qtempdepsol}), we propose the following fit for the amplitude,
 \begin{eqnarray}
 Q_1 (z,t_1) &=& Q_0(z) \left[ \frac{\left( L - \lp \right) \left( L + \frac{1}{3} \lp \right) - 4 \left( z_0 + t_1 L \right)^2}{\left( L - \lp \right) \left( L + \frac{1}{3} \lp \right) - 4 z_0^2} \right]^{n},\label{eq:fit}
\end{eqnarray}
with $n$ an empirical exponent. When Equation~(\ref{eq:fit}) is applied to the results of Figure~\ref{fig:flow1}(a), we obtain that the exponent $n = 1$ provides a good fit for the amplitude (see the dotted line in Figure~\ref{fig:flow1}(a)).

On the other hand, we perform in Figure~\ref{fig:flow1}(b) a wavelet power spectrum \citep{wavelet} of the signal displayed in Figure~\ref{fig:flow1}(a). We find that the period decreases in time as the thread moves towards the footpoint of the magnetic structure. The WKB approximation for the period given by Equation~(\ref{eq:period}) is in excellent agreement with the position of the maximum of the wavelet spectrum (see dashed line in Figure~\ref{fig:flow1}(b)). As discussed in Section~\ref{sec:amplitude}, the WKB approximation for the period is much more accurate than the WKB approximation for the amplitude. In addition, we see that the rate at which the period changes with respect to the value at $t=0$ is not constant. Using  Equation~(\ref{eq:period}) and considering the parameters of this particular simulation, the period of the oscillation when the thread is at $z=L/4$ has decreased of about 9\% with the respect to its initial value at $z=0$, whereas the period decreases of about 45\% when the thread finally reaches the footpoint of the magnetic tube at $z=L$.

We repeat the simulation for $\lp/L = 0.2$ and a faster flow, $v_0 / \vap = 0.1$, and take  $z_0 / L = -0.25$ to consider  the thread initially displaced from the center of the flux tube. The radial velocity perturbation at $z=0$ is plotted in Figure~\ref{fig:flow1}(c), and the corresponding wavelet power spectrum is shown in Figure~\ref{fig:flow1}(d). These results are equivalent to those of  Figures~\ref{fig:flow1}(a)--(b), i.e., both the amplitude and the period of the oscillation depend on the position of the prominence thread within the flux tube, taking both of them their maximum value when the thread is centered. As before, Equation~(\ref{eq:period}) is a very good approximation to the period.

\subsection{Arbitrary excitation}

\begin{figure*}[!t]
\centering
 \includegraphics[width=0.79\columnwidth]{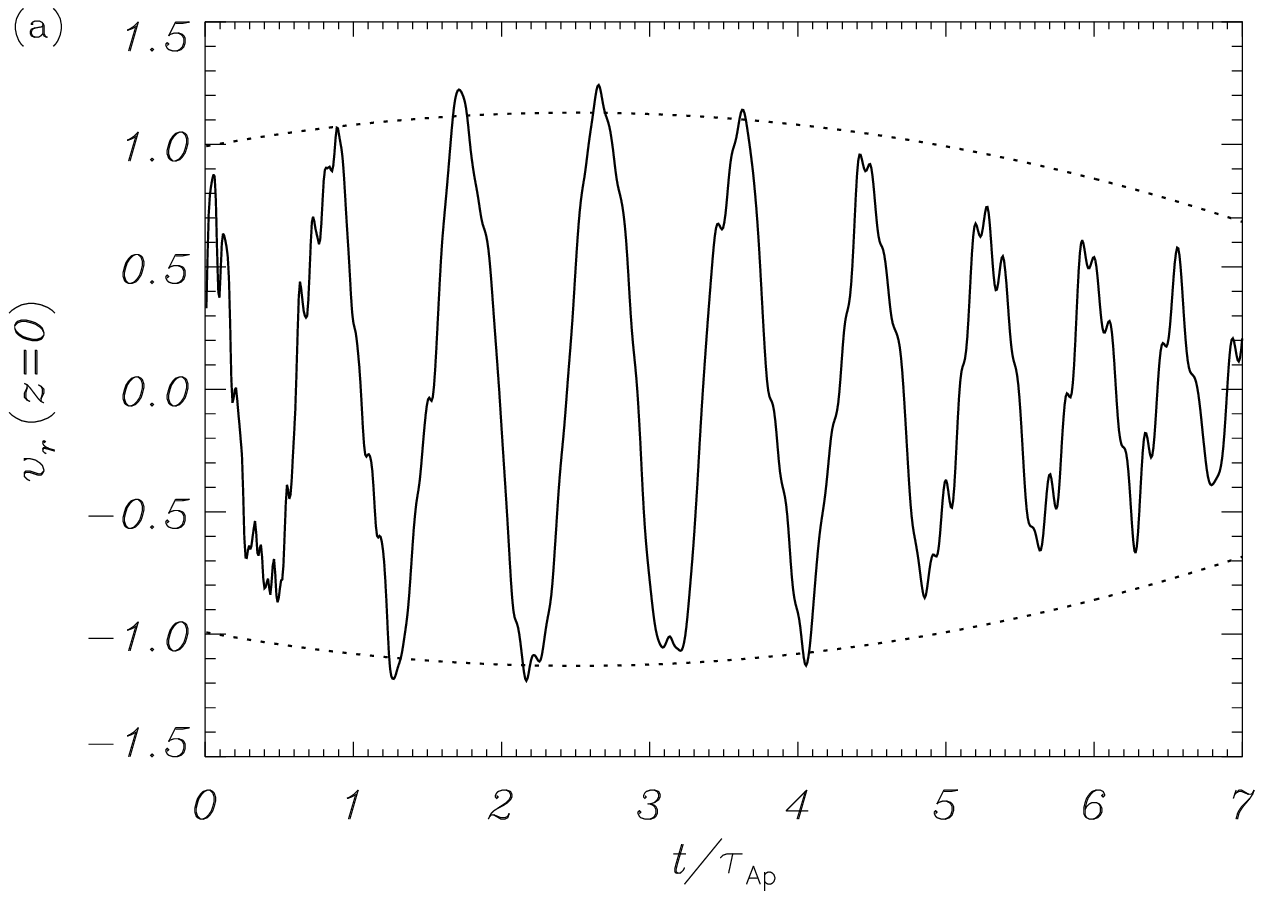}
 \includegraphics[width=0.79\columnwidth]{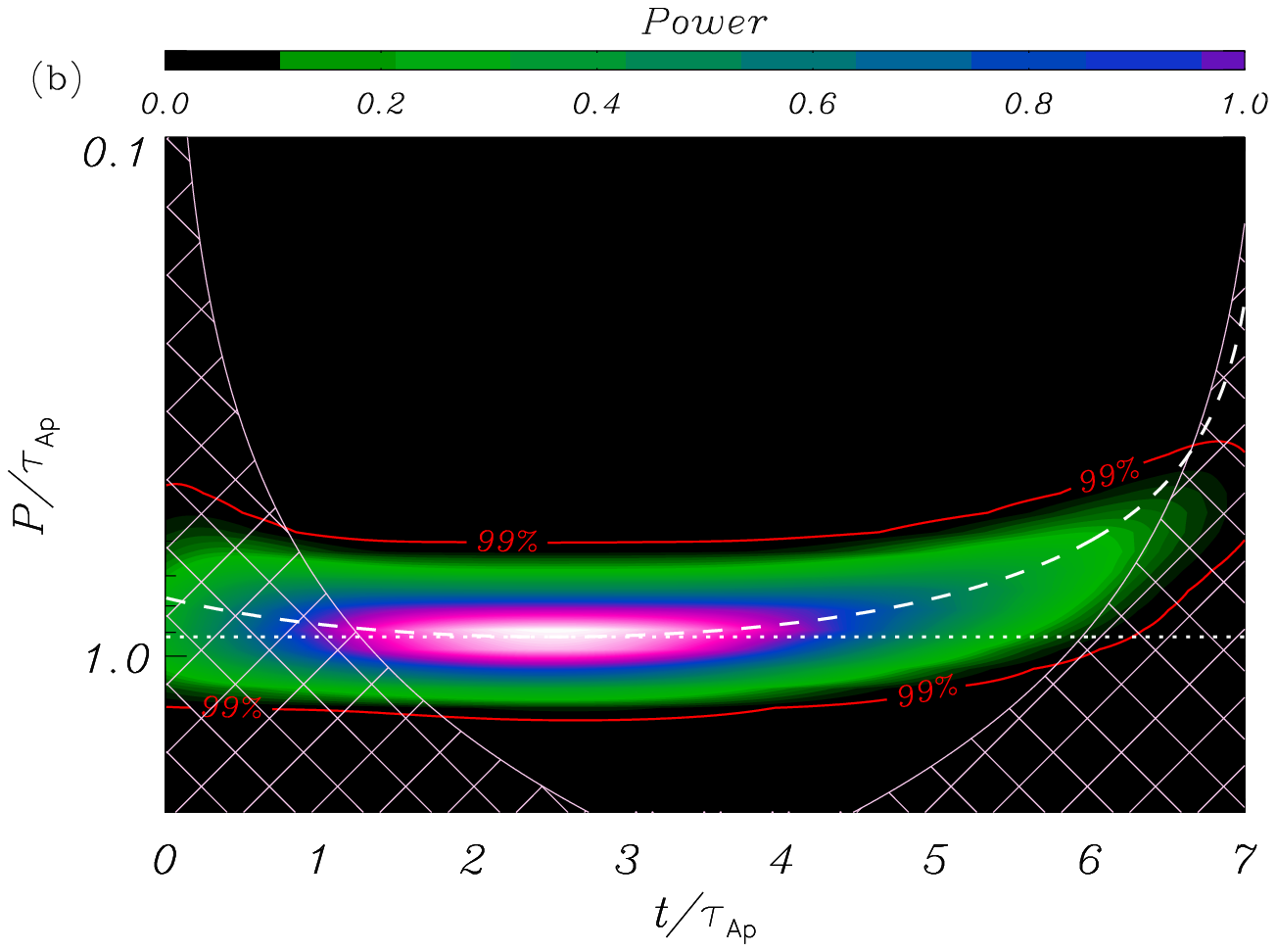} 
\includegraphics[width=0.79\columnwidth]{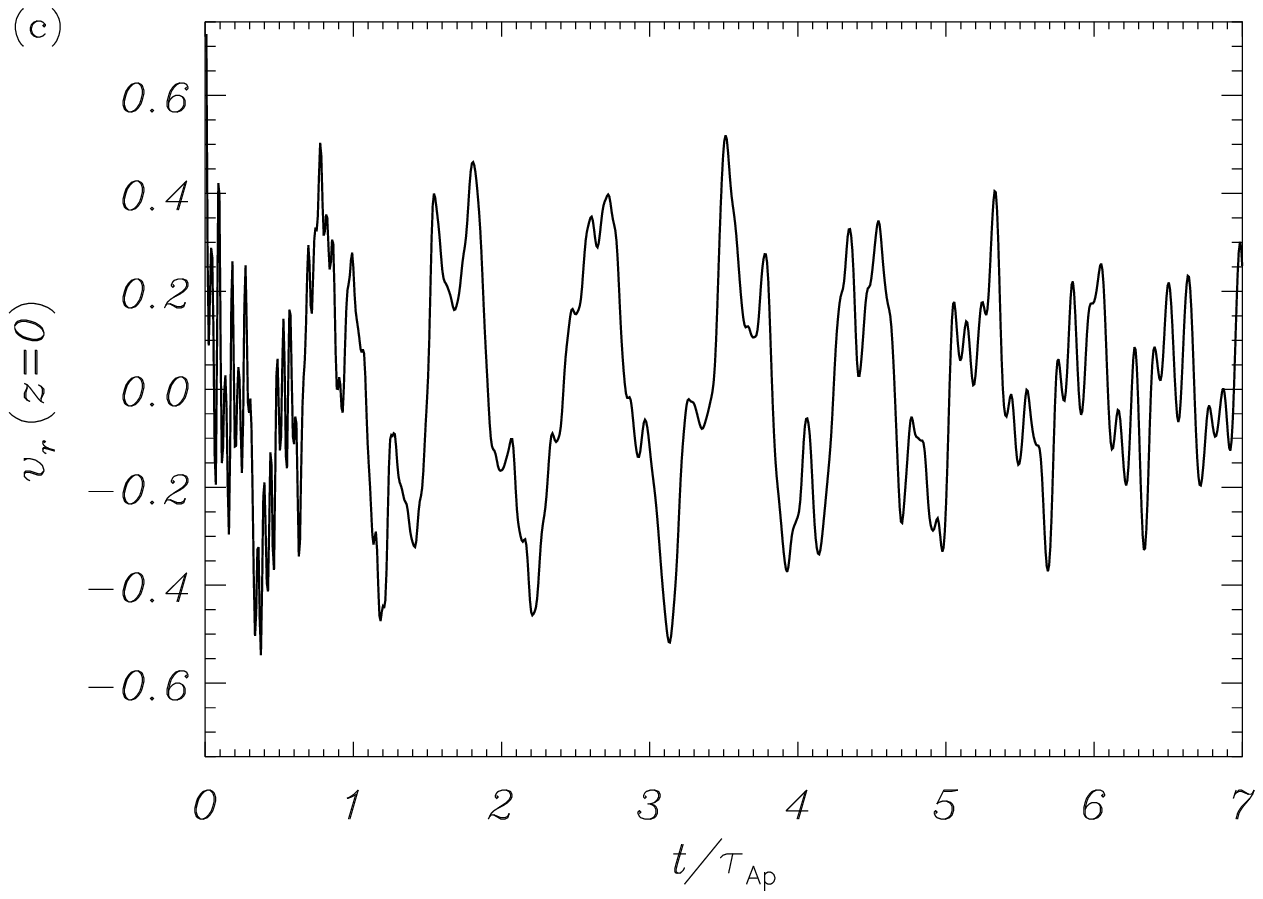}
 \includegraphics[width=0.79\columnwidth]{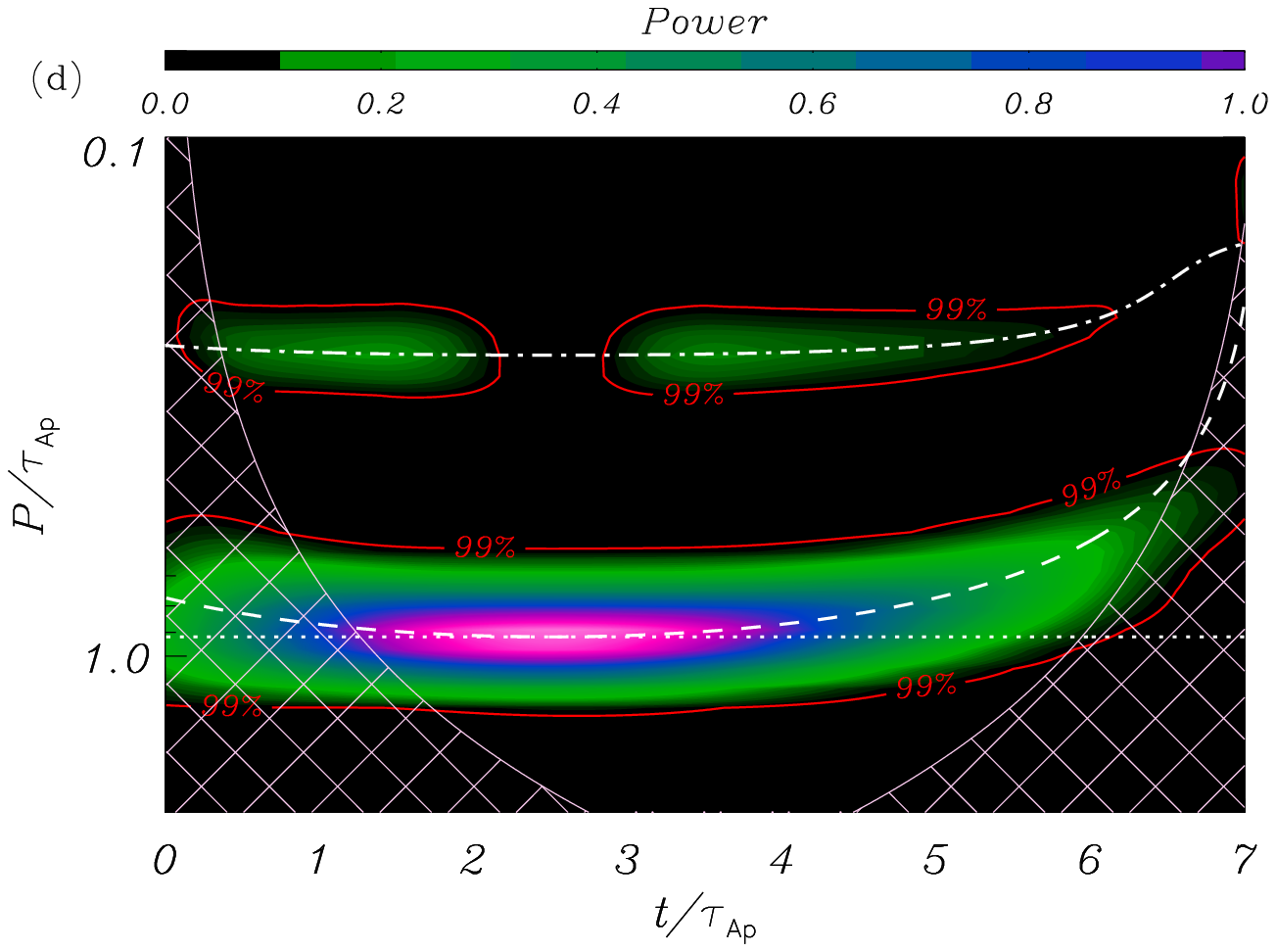}
\caption{(a) Same as Figure~\ref{fig:flow1}(c) but for a initial excitation given by Equation~(\ref{eq:excitat}) with $\zeta / L = -0.25 $ and $\sigma / L = 0.2$. (b) Wavelet power spectrum corresponding to the signal displayed in panel (a). (c) Same as panel (a) but for $\zeta / L = 0 $ and $\sigma / L = 0.2$. (d) Same as panel (b) but for the signal displayed in panel (c), with the dash-dotted line denoting the period of the first harmonic. \label{fig:arb}}
\end{figure*}

%

In the previous Section, we have used an initial condition for $v_r$ that corresponds to the fundamental mode eigenfunction, so only this mode is excited. However, it is expected that the energy from an arbitrary disturbance of the flux tube is deposited in many normal modes \citep[see a discussion on this issue in, e.g.,][]{terradasexc}.  To represent an arbitrary disturbance of the flux tube, we consider a Gaussian function as the initial condition of $v_r$ at $t=0$, namely 
\begin{equation}
 v_r (z,t=0) = \exp \left[ - \frac{\left( z - \zeta \right)^2}{\sigma^2}  \right], \label{eq:excitat}
\end{equation}
 where $\zeta$ and $\sigma$ are arbitrary parameters. Whereas $\zeta$ correspond to the position of the maximum of the excitation, $\sigma$ determines its width.

In the following simulations, we consider the same model parameters as in Figures~\ref{fig:flow1}(c)--(d), i.e., $\lp / L = 0.2$, $v_0 / \vap = 0.1$, and $z_0 / L = -0.25$, but use the initial condition given by Equation~(\ref{eq:excitat}). To begin with, we take $\sigma/L = 0.2$ and consider different values of $\zeta$. First we use $\zeta / L = -0.25$, so the excitation is mainly confined to the dense prominence region of the flux tube. The result of this simulation is displayed in Figure~\ref{fig:arb}(a), which shows the evolution in time of $v_r$ at $z=0$, whereas Figure~\ref{fig:arb}(b) shows the corresponding wavelet power spectrum. It is interesting to compare Figures~\ref{fig:flow1}(d) and \ref{fig:arb}(b) to see that, in the present case, the oscillation dynamics is still governed by the fundamental normal mode. We see in  Figure~\ref{fig:arb}(a) that there is some contribution of higher harmonics to the behavior of $v_r$ in time, although their contribution to the overall oscillation is of very minor importance. In addition, the evolution of the amplitude of $v_r$ remains qualitatively described by Equation~(\ref{eq:fit}) with $n=1$.

Next, we perform another simulation by taking the same parameters as before but assuming $\zeta / L = 0$. In this case, the maximum of the initial excitation is located in the evacuated part of the magnetic tube. Again, we plot in Figure~\ref{fig:arb}(c) $v_r$ at $z=0$ versus time, and in Figure~\ref{fig:arb}(d) the wavelet power spectrum. The behavior of $v_r$ in time is substantially different in the present situation compared to the case of Figures~\ref{fig:arb}(a)--(b). First of all, we see that $v_r$ is not governed by the fundamental mode exclusively. The wavelet power spectrum indicates that the energy from the initial disturbance is mainly deposited to the fundamental mode, but also the first harmonic is excited. The dependence in time of the fundamental mode period is again well described by Equation~(\ref{eq:period}). However, the contribution of the first harmonic to the overall oscillation seems to have disappeared when the thread is located at the center of the magnetic tube. The reason for this result is that the first harmonic eigenfunction has a node at $z=0$ when the thread is centered, and so the first harmonic does not contribute to the signal displayed in Figure~\ref{fig:arb}(c) in such a case. On the other hand, it is now difficult to determine the effect of the flow on the amplitude of the oscillation. 

Finally, we have performed several simulations for different values of $\sigma$. If the maximum of the excitation is located within the dense part of the flow tube, the results are rather insensitive to $\sigma$ unless values much smaller than $\lp$ are used. In all the cases, the fundamental mode is predominantly excited. However, the results are more affected by the value of $\sigma$ if the maximum of the excitation is located in the evacuated part of the tube. In such a case, the larger $\sigma$, the more energy is deposited in the fundamental mode. On the contrary, as $\sigma$ gets smaller the energy of the initial excitation is more distributed among higher harmonics. 

The results of this Section point out that the time-dependent behavior of standing kink MHD waves of flowing prominence threads is strongly influenced by the form of the initial  disturbance. If the initial disturbance mainly perturbs the dense prominence part of the flux tube, the oscillations are governed by the fundamental kink mode. In such a case, the dependence of both the period and the amplitude with the flow velocity are approximately given by Equations~(\ref{eq:period}) and (\ref{eq:fit}), respectively.  On the contrary, the behavior is more complicated if the initial perturbation takes place in the evacuated part of the fine structure as other harmonics are excited in addition to the fundamental mode. The contribution of the different harmonics depends on both the position  and the width of the initial excitation, while the amplitude of the oscillation does not have a simple dependence on the flow velocity.

\section{Discussion and conclusions}
\label{sec:discussion}

In this paper, we have investigated standing kink MHD waves in the fine structure of solar prominences, modeled as coronal magnetic flux tubes partially filled with flowing threads of prominence material. The present study extends and complements the previous work by \citet{hinode}, who restricted themselves to the numerical investigation of this phenomenon and did not perform an in-depth parametric study. Here, we have combined analytical methods based on the WKB approximation with time-dependent numerical simulations to assess the precise effect of the flow on both the period and the amplitude of the fundamental kink mode.

As for the effect of the flow on the period, we can distinguish two different situations. On the one hand, we find that the flow has a small effect on the period when the thread is located near the center of the supporting magnetic flux tube. In this case, the variation of the period with respect to the static case may fall within the error bars of the observations, and so the effect may be undetectable. There our results confirm the qualitative discussion of \citet{hinode} about the effect of the flow on the period. On the other hand, the variation of the period is much more important when the thread approaches the footpoint of the magnetic structure. Then, the decrease of the period can be larger than 50\% with respect to the static case. The case in which the thread is near one of the footpoints of the magnetic tube was not analyzed by \citet{hinode}. 

We have also found that the flow affects the amplitude of the fundamental mode. This result was not discussed by \citet{hinode}. During the motion of the prominence thread along the magnetic structure, we find that the amplitude grows as the thread gets closer to the center of the tube and decreases otherwise. This produces an apparent amplification or damping of the oscillations, respectively. Observations often indicate that thread transverse oscillations are strongly damped \citep[see, e.g.,][]{lin04,linrev,ning}. While several mechanisms have been proposed and investigated to explain the quick attenuation \citep[see the recent reviews by][]{oliver,arreguiballester}, the process of resonant absorption seems the most likely explanation \citep[e.g.,][]{arregui08,solerslow,solerRAPI,solerstatic}. Our present results indicate that the actual damping rate of the oscillations might be affected by the change of the amplitude due to the flow. This fact should be taken into account when the damping rate is used as a seismological tool to infer physical parameters of prominence threads, because the presence of flow may introduce some uncertainties on these estimations \citep[see details in][]{arreguiballester}.

In addition, our numerical simulations have allowed us to determine how different perturbations excite the oscillations of the magnetic structure. Based on the cases studied in this paper, we have obtained that the fundamental mode is mostly excited when the perturbation initially disturbs the dense, prominence part of the tube. From the wavelet power spectrum of the radial velocity perturbation, we conclude that the contribution of higher harmonics is negligible, thus the overall oscillation is governed by the fundamental mode. On the contrary, a perturbation located at the evacuated part of the tube excites the fundamental mode and higher harmonics, producing a more complex behavior of the oscillations. In this last case, the effect of the flow on the amplitude is more complicated and no simple dependence can be extracted from the simulations. 

This paper has explored the properties of MHD waves in a coronal magnetic structure with a changing configuration. Previous similar works in this line are, e.g., \citet{hinode} in prominences, and \citet{morton1,morton2,morton3} in coronal loops. During the revision of this paper it also came to our knowledge the recent work by \citet{ruderman}. In view of the highly dynamic nature of the coronal medium in general, and the prominence plasma in particular, this kind of modeling represents a better description of the actual oscillatory phenomena in the corona and in prominences. The present investigation could be extended in the future by incorporating the effect of the density inhomogeneity in the transverse direction and so investigating the resonant damping of the kink mode. 

\acknowledgements{
   RS thanks J. L. Ballester, R. Oliver, and T. Van Doorsselaere for useful comments. RS acknowledges support from a Marie Curie Intra-European Fellowship within the European Commission 7th Framework Program (PIEF-GA-2010-274716). RS also  thanks support from the EU Research and Training Network ``SOLAIRE'' (MTRN-CT-2006-035484). RS acknowledges discussion within ISSI Team on ``Solar Prominence Formation and Equilibrium: New data, new models'', and is grateful to ISSI for the financial support. MG acknowledges support from K.U. Leuven via GOA/2009-009. Wavelet software was provided by C. Torrence and G. Compo, and is available at http://paos.colorado.edu/research/wavelets/}



\begin{thebibliography}{}
%
 \bibitem[Ahn et al.(2010)]{ahn2010} Ahn, K., Chae, J.,  Cao, W., \& Goode, P. R. 2010, \apj, 721, 74

\bibitem[Andries et al.(2005a)]{andries2005a} Andries, J., Goossens, M., Hollweg, J. V., Arregui, I., \& Van Doorsselaere, T. 2005a, \aap, 430, 1109

\bibitem[Andries et al.(2005b)]{andries2005b} Andries, J., Arregui, I., \& Goossens, M. 2005b, \apj, 624, L57

\bibitem[Andries et al.(2009)]{andriesrev} Andries, J., Van Doorsselaere, T., Roberts, B., Verth, G., Verwichte, E., \& Erd\'elyi, R. 2009, \ssr, 149, 3

\bibitem[Antolin \& Verwichte(2011)]{antolin} Antolin, P., \& Verwichte, E. 2011, \apj, in press (arXiv:1105.2175)

  \bibitem[Arregui et al.(2008)]{arregui08} Arregui, I., Terradas, J., Oliver, R., \& Ballester, J. L. 2008, \apj, 682, L141 



\bibitem[Arregui \& Ballester(2010)]{arreguiballester} Arregui, I., \& Ballester, J. L. 2010, \ssr, in press (DOI:10.1007/s11214-010-9648-9) 


\bibitem[Ballester(2006)]{ballester} Ballester, J. L. 2006, Phil. Trans. R. Soc. A, 364, 405 


\bibitem[Bender \& Orszag(1978)]{bender} Bender, C. M., \& Orszag, S. A. 1978, Advanced Mathematical Methods for Scientists and Engineers (New York: McGraw-Hill)

 \bibitem[Berger et al.(2008)]{berger} Berger et al. 2008, \apj, 676, L89

 \bibitem[Cao et al.(2010)]{cao2010} Cao, W., Ning, Z., Goode, P. R., Yurchyshyn, V., \& Haisheng, J. 2010, \apj, 719, L95


 \bibitem[Carbonell \& Ballester(1991)]{carball} Carbonell, M., \& Ballester, J. L. 1991, \aap, 249, 295


 \bibitem[Chae et al.(2008)]{chae} Chae, J., Ahn, K., Lim, E.-K., Choe, G. S., \& Sakurai, T. 2008, \apj, 689, L73

 \bibitem[Chae(2010)]{chae2010} Chae, J. 2010, \apj, 714, 618

\bibitem[De Pontieu \& McIntosh(2010)]{depontieu2010} De Pontieu, B., \& McIntosh, S. W. 2010, \apj, 722, 1013

 \bibitem[D\'\i az et al.(2001)]{diaz2001} D\'iaz, A. J., Oliver R., Erd\'elyi, R., \& Ballester, J. L.2001, \aap, 379, 1083

 \bibitem[D\'\i az et al.(2002)]{diaz2002} D\'iaz, A. J., Oliver R., \& Ballester, J. L. 2002, \apj, 580, 550

 \bibitem[D\'\i az et al.(2010)]{diazperiods} D\'iaz, A. J., Oliver R., \& Ballester, J. L. 2010, \apj, 725, 1742

\bibitem[Dymova \& Ruderman(2005)]{dymovaruderman} Dymova, M. V., \& Ruderman, M. S. 2005, \solphys, 229, 79


\bibitem[Edwin \& Roberts(1983)]{edwinroberts} Edwin, P. M., \& Roberts, B. 1983, \solphys, 88, 179



\bibitem[Erd\'elyi \& Fedun(2007)]{erdelyi2007} Erd\'elyi, R., \& Fedun, V. 2007, Science, 318, 1572

 \bibitem[Goedbloed \& Poedts(2004)]{goedbloed} Goedbloed, H., \& Poedts, S. 2004, Principles of magnetohydrodynamics, Cambridge University Press

\bibitem[Goossens et al.(2009)]{goossens2009} Goossens, M., Terradas, J., Andries, J., Arregui, I., Ballester, J. L. 2009, \aap, 503, 213

\bibitem[Jess et al.(2009)]{jess2009} Jess, D. B., et al. 2009, Science, 323, 1582

  \bibitem[Joarder \& Roberts(1992)]{joarder} Joarder, P. S. \& Roberts, B. 1992, \aap, 261, 625

\bibitem[Joarder et al.(1997)]{joarder97} Joarder, P. S., Nakariakov, V. M., \& Roberts, B. 1997, \solphys, 173, 81


\bibitem[Kucera et al.(2003)]{kucera2003} Kucera, T. A., Tovar, M., \& De Pontieu, B. 2003, \solphys, 212, 81

 \bibitem[Lin et al.(2003)]{lin2003} Lin, Y., Engvold, O., \& Wiik, J. E. 2003, \solphys, 216, 109

   \bibitem[Lin(2004)]{lin04} Lin, Y. 2004, PhD Thesis, University of Oslo, Norway

 \bibitem[Lin et al.(2008)]{lin08} Lin, Y., Martin, S. F., \& Engvold, O. 2008, in ASP Conf. Ser. 383, Subsurface and Atmospheric Influences on Solar Activity, ed. R. Howe, R. W. Komm, K. S. Balasubramaniam, \& G. J. D. Petrie (San Francisco: ASP), 235
 
\bibitem[Lin et al.(2009)]{lin09} Lin, Y., Soler, R., Engvold, O., Ballester, J. L., Langangen, \O., Oliver, R., \& Rouppe van der Voort, L. H. M. 2009, \apj, 704, 870

\bibitem[Lin(2010)]{linrev} Lin, Y. 2010, \ssr, in press (DOI:10.1007/s11214-010-9672-9)

 \bibitem[Morton et al.(2010)]{morton1} Morton, R. J., Hood, A. W., \& Erd\'elyi, R. 2010, \aap, 512, A23
\bibitem[Morton \& Erd\'elyi(2010a)]{morton2} Morton, R. J., \& Erd\'elyi, R. 2010a, \apj, 707, 750
\bibitem[Morton \& Erd\'elyi(2010b)]{morton3} Morton, R. J., \& Erd\'elyi, R. 2010b, \aap, 519, A43

 \bibitem[Nakariakov \& Roberts(1995)]{nakaroberts} Nakariakov, V. M., \& Roberts, B. 1995, \solphys, 159, 213


 \bibitem[Ning et al.(2009)]{ning} Ning, Z., Cao, W., Okamoto, T. J., Ichimoto, K., \& Qu, Z. Q. 2009, \aap, 499, 595

\bibitem[Ofman \& Wang(2008)]{ofmanwang} Ofman, L., \& Wang, T. J. 2008, \aap, 482, L9 

 \bibitem[Okamoto et al.(2007)]{okamoto} Okamoto, T. J, et al. 2007, Science, 318, 1557 

  \bibitem[Oliver et al.(1993)]{oliverslab} Oliver, R., Ballester, J. L., Hood, A. W., \& Priest, E. R. 1993, \apj, 409, 809

 \bibitem[Oliver(2009)]{oliver} Oliver, R. 2009, \ssr, 149, 175

 \bibitem[Ruderman(2011)]{ruderman} Ruderman, M. S. 2011, \solphys, in press (DOI:10.1007/s11207-011-9772-z)


\bibitem[Schmieder et al.(2010)]{brigi} Schmieder, B., Chandra, R., Berlicki, A., \& Mein, P. 2010, \aap, 514, A68

 \bibitem[Sewell(2005)]{sewell} Sewell, G. 2005, The Numerical Solution of Ordinary and Partial Differential Equations (Hoboken: Wiley \& Sons)
 
   \bibitem[Soler et al.(2009a)]{solerslow} Soler, R., Oliver, R., Ballester, J. L., \& Goossens, M. 2009a, \apj, 695, L166

\bibitem[Soler et al.(2009b)]{solerRAPI} Soler, R., Oliver, R., \& Ballester, J. L. 2009b,  \apj, 707, 662



\bibitem[Soler et al.(2010)]{solerstatic} Soler, R., Arregui, I., Oliver, R., \& Ballester, J. L. 2010,  \apj, 722, 1778

 \bibitem[Terradas et al.(2007)]{terradasexc} Terradas, J., Andries, J., \& Goossens. M. 2007, \aap, 469, 1135

 \bibitem[Terradas et al.(2008)]{hinode} Terradas, J., Arregui, I., Oliver, R., \& Ballester, J. L. 2008, \apj, 678, L153

 \bibitem[Terradas et al.(2010)]{TGV} Terradas, J., Goossens, M., \& Verth, G. 2010, \aap, 524, A23

 \bibitem[Terradas et al.(2011)]{terradasflow} Terradas, J., Arregui, I., Verth, G., \& Goossens, M. 2011, \apj, 729, L22

 \bibitem[Terra-Homem et al.(2003)]{terra} Terra-Homem, M., Erd\'elyi, R., \& Ballai, I. 2003, \solphys, 217, 199


 \bibitem[Tomczyk et al.(2007)]{tomczyk2007} Tomczyk, S., et al. 2007, Science, 317, 1197

\bibitem[Tomczyk \& McIntosh(2009)]{tomczyk2009} Tomczyk, S., \& McIntosh, S. W. 2010, \apj, 697, 1384


\bibitem[Torrence \& Compo(1998)]{wavelet} Torrence, C., \& Compo, G. P. 1998, Bull. Amer. Meteor. Soc., 79, 61

 \bibitem[Van Doorsselaere et al.(2008)]{tom2008} Van Doorsselaere, T., Nakariakov, V. M., \& Verwichte, E. 2008, \apj, 676, L73

 \bibitem[Verth et al.(2010)]{VTG2010} Verth, G., Terradas, J., \& Goossens, M. 2010, \apj, 718, L102

\bibitem[Wang(1999)]{wang1999} Wang, Y.-M. 1999, \apj, 520, L71

 \bibitem[Wang et al.(2009)]{wang2009} Wang, T. J., Ofman, L., Davila, J. M., \& Mariska, J. T. 2009, \aap, 503, L25

 \bibitem[Zirker et al.(1998)]{zirker98} Zirker, J. B., Engvold, O., \& Martin, S. F. 1998, Nature, 396, 440
%
%
   \end{thebibliography}
\end{document}